\documentclass[12pt]{article}
\usepackage{graphicx}
\usepackage{subcaption}
\usepackage{amsmath,amssymb,xcolor,physics}
\usepackage{url}
\usepackage{ulem}
\usepackage{hyperref}
\usepackage{tabularray}
\usepackage{tikz}
\usetikzlibrary{shapes.geometric, arrows, positioning, calc}
\tikzstyle{theory} = [rectangle,  draw=black, minimum width=3cm, minimum height=0.6cm, text centered, fill=gray!10]
\tikzstyle{arrow} = [thick,->,>=stealth]
\tikzstyle{whitelabel} = [fill=white, text centered, inner sep=2pt]

\allowdisplaybreaks[4]
\usepackage{float}

\begin{document}

\begin{center}

{\Large
The Kogut-Susskind Hamiltonian from the orbifold lattice }\\
\vspace{0.2cm}
{\Large
at finite lattice spacing}

\end{center}

\vspace{0.2cm}
\begin{center}
Georg Bergner$^{\rm a}$, Masanori Hanada$^{\rm b}$, Emanuele Mendicelli$\, ^{\rm c}$
\end{center}

\vspace{0.3cm}
\begin{center}
$^{\rm a}$
Institute for Theoretical Physics, University of Jena\\
Max-Wien-Platz 1, 07743 Jena, Germany\\
\vspace{1mm}
$^{\rm a}$
Leibniz Institute of Photonic Technology, Albert-Einstein-Str. 9, 07745, Jena, Germany\\
\vspace{1mm}
$^{\rm b}$
School of Mathematical Sciences, Queen Mary University of London\\
Mile End Road, London, E1 4NS, United Kingdom\\
\vspace{1mm}
$^{\rm b}$
qBraid Co., Harper Court 5235, Chicago, IL 60615, United States\\
\vspace{1mm}
$^{\rm c}$\, Department of Mathematical Sciences, University of Liverpool\\
Liverpool L69 7ZL, United Kingdom\\
\end{center}

\vspace{0.5cm}

\begin{center}
  {\bf Abstract}
\end{center}
We demonstrate that the orbifold lattice Hamiltonian -- an approach known for its efficiency in simulating SU($N$) Yang-Mills theory and QCD on digital quantum computers -- can reproduce the Kogut-Susskind Hamiltonian in a controlled limit. We show that it emerges naturally as the infinite scalar mass limit of the orbifold lattice formulation, even at finite lattice spacing. Our analysis provides both a general analytical framework applicable to SU($N$) gauge theories in arbitrary dimensions and specific numerical evidence for $(2+1)$-dimensional SU($N$) Yang-Mills theories ($N=2,3$). Using Euclidean path integral methods, we quantify the approach to this limit by comparing the standard Wilson action with the orbifold lattice action, matching lattice parameters, and systematically extrapolating results as the bare scalar mass approaches infinity. 
Because the orbifold lattice and the Kogut-Susskind formulation share the same continuum limit, this limit does not have to be taken in practice: it is an option rather than a requirement. Should one wish to take it -- for instance to reproduce Kogut-Susskind results at finite lattice spacing, including their lattice artifacts -- we find that a bare scalar mass of order of the inverse lattice spacing already suffices, so that no energy scale beyond the cutoff is introduced. The orbifold lattice Hamiltonian admits quantum simulation protocols whose gate cost per Trotter step, and whose classical compilation cost, grow only polynomially with the number of qubits $Q$ assigned to each bosonic degree of freedom -- exponentially better in $Q$ than protocols whose cost grows as a power of the truncation level $\Lambda = 2^Q$ -- for any gauge group and any spatial dimension. The equivalence established here allows this property to be inherited by the Kogut-Susskind Hamiltonian. We emphasize that the present work establishes the equivalence using classical Euclidean Monte Carlo; the speedup itself is a property of the quantum simulation of the orbifold Hamiltonian established in earlier work.

\newpage
\tableofcontents
%%%%%%%%%%%%%%%%%%%%%%%
%%%%%%%%%%%%%%%%%%%%%%%
\section{Introduction}
%%%%%%%%%%%%%%%%%%%%%%%
%%%%%%%%%%%%%%%%%%%%%%%
The Kogut-Susskind Hamiltonian~\cite{Kogut:1974ag} has emerged as a canonical framework for quantum simulation of Yang-Mills theory and QCD on quantum computing architectures. 
However, this approach presents a fundamental incongruity: the formalism, conceived prior to the conceptualization of quantum computation~\cite{Feynman:1981tf}, was inherently not designed with quantum algorithmic efficiency in mind. Its reliance on compact variables --- specifically unitary link variables --- engenders a Hilbert space whose structural complexity becomes increasingly intractable for SU($N\ge 2$) when extended beyond one spatial dimension, particularly when the continuum limit is considered. This intrinsic limitation has manifested as a persistent obstacle; despite two decades of scholarly pursuit since the inaugural quantum simulation protocol~\cite{Byrnes:2005qx}, the field has yet to produce a concrete implementation capable of demonstrating genuine quantum advantage in the continuum limit~\cite{Hanada:2025yzx}.
Although simulations on quantum hardware (digital or analog), including refs.~\cite{Martinez:2016yna,Yang:2020yer,Klco:2019evd, ARahman:2021ktn, ARahman:2022tkr, Atas:2021ext, Ciavarella:2024lsp, Kavaki:2024ijd,Than:2024zaj,Ciavarella:2021nmj, Illa:2022jqb, Ciavarella:2021lel, Atas:2022dqm, Farrell:2022wyt, Farrell:2022vyh, Ciavarella:2023mfc, Ciavarella:2024fzw, Hayata:2024fnh}, have led to promising results for certain limited cases, they leveraged specific features of systems (e.g., Abelian, low dimensions, or truncation to small number of levels) which cannot be extended to SU($N$) in higher spatial dimensions.

In any digital simulation, each degree of freedom of the theory has to be encoded into a finite number of qubits, and the cost is naturally measured in terms of this number. We write $Q$ for the number of qubits assigned to each bosonic degree of freedom, so that the dimension of the local Hilbert space is $\Lambda=2^{Q}$. Throughout, ``exponential'' refers to the scaling with $Q$, at the level of the gate cost of a single Trotter step and of the classical cost of compiling that step. The required value of $Q$ is comparable in the two formulations: in both, the truncation error decays exponentially in $\Lambda$, so that $Q=O(\log\log 1/\epsilon)$ suffices for a target accuracy $\epsilon$~\cite{Tong:2021rfv,Hanada:2025yzx}. The number of Trotter steps is controlled by commutator norms taken over the truncated Hilbert space, and these grow with the cutoff in either formulation. We therefore make no claim regarding the number of Trotter steps; the two formulations stand on the same footing in this respect, and our comparison is made at the level of the per-step cost.
The fundamental constraints are twofold: existing protocols necessitate classical preprocessing with computational demands that scale exponentially with qubit count, while the quantum circuit depth likewise exhibits exponential scaling relative to the number of qubits allocated to each bosonic degree of freedom. Because of this, no quantum simulation protocol for the Kogut-Susskind Hamiltonian is known whose cost toward the continuum limit is under control in both its scaling with the truncation and its prefactor~\cite{Hanada:2025yzx}.

The fundamental constraints are twofold: existing protocols necessitate classical preprocessing with computational demands that scale exponentially with qubit count, while the quantum circuit depth likewise exhibits exponential scaling relative to the number of qubits allocated to each bosonic degree of freedom. Because of this, there is no known quantum simulation protocol that allows us to take the continuum limit of the Kogut-Susskind Hamiltonian~\cite{Hanada:2025yzx}; the readers are encouraged to check the state-of-the-art effort on the explicit circuit construction of the SU(3) theory on three spatial dimensions~\cite{Balaji:2025afl} that observed significant resource requirement.
Strictly speaking, even asymptotic gate counting is difficult for the Kogut-Susskind approach: such counts have been obtained only for U(1), SU(2) and SU(3), through a separate derivation for each group~\cite{Kan:2021xfc}, and no count is available for general SU($N$).
In the early protocol of ref.~\cite{Byrnes:2005qx}, the magnetic evolution requires $\tilde{O}(\Lambda^{4})$, $\tilde{O}(\Lambda^{12})$ and $\tilde{O}(\Lambda^{32})$ T gates per Trotter step for U(1), SU(2) and SU(3), respectively~\cite[Table~II]{Kan:2021xfc}; with $\Lambda=2^{Q}$ this is exponential in $Q$.
Ref.~\cite{Kan:2021xfc} reduced this to $\tilde{O}(\log^{2}\Lambda)$ T gates per Trotter step for both the electric and the magnetic terms, for U(1), SU(2) and SU(3) in arbitrary spatial dimension.
For the smallest three-dimensional example considered there --- transport coefficients in SU(3) with $\Lambda=10$, $L=10$, $a=0.1$, $T=1$ and $\epsilon=10^{-8}$ --- the estimate is $3.0\times10^{49}$ T gates~\cite[Sec.~V\,C]{Kan:2021xfc}; the prefactor responsible is independent of $\Lambda$ and of the qubit count.
Ref.~\cite{Rhodes:2024zbr} provided a counting that does not include the cost of implementing the sparse-access oracles; an explicit construction of these oracles for the SU(3) magnetic term is not given there~\cite{Draper:2026bcj}, and the block-encoding normalisation may depend on the sparsity of the plaquette operator in a way not reflected in the quoted gate costs~\cite{Davoudi:2026pcu}. 
Whether these oracles can be implemented without significant additional cost remains to be studied~\cite{Hanada:2025yzx}.
For the SU(3) theory in $3+1$ dimensions, no construction based on the Kogut-Susskind Hamiltonian has yet been demonstrated that combines a gate cost polynomial in $Q$ with a prefactor that would permit execution.\footnote{After the first version of this paper appeared, two relevant papers were published. The authors of ref.~\cite{Davoudi:2022xmb} applied their Hamiltonian-splitting strategy to the SU(3) Kogut-Susskind Hamiltonian in arbitrary spatial dimension~\cite{Davoudi:2026pcu}, reducing the number of subevolutions per plaquette and Trotter step from $\sim1.5\times10^{15}$~\cite{Kan:2021xfc} to $\sim1.1\times10^{6}$, and the T-gate estimate by a factor of nearly $10^{14}$ independently of simulation parameters, which for the example quoted above corresponds to roughly $10^{36}$ T gates. The scaling with $\Lambda$, and hence with $Q$, is unchanged, and the implementation of the individual subevolutions is deferred to refs.~\cite{Kan:2021xfc,Davoudi:2022xmb}. Ref.~\cite{Draper:2026bcj} gave an explicit, compiled block encoding of the SU(3) plaquette operator in three spatial dimensions in the gauge-invariant irrep basis, for truncations keeping the link irreps $\{\mathbf{1},\mathbf{3},\bar{\mathbf{3}},\mathbf{8},\mathbf{6},\bar{\mathbf{6}}\}$, at $10^{5}$--$10^{6}$ T gates per call. The construction is built from classically precomputed lookup tables whose size grows with the number of site states, i.e.\ polynomially in the cutoff and hence exponentially in $Q$. The circuit is not given in closed form as a function of the truncation: the tables are regenerated from SU(3) Clebsch-Gordan data for each truncation (the author quotes a growth of the corner tables by a factor of about 300 between the truncations $B=9$ and $B=16$), and the full time evolution scales as $L^{6}$ with the linear lattice size. Thus a gate cost polynomial in $Q$ has so far been achieved only with prefactors of order $10^{36}$--$10^{49}$, while the constructions with manageable prefactors have a cost that grows exponentially in $Q$.}
In most cases, even a resource estimate is difficult; see ref.~\cite{Hanada:2025yzx,Hanada:2026zab} for details.

To provide a viable way to realize quantum simulations of SU($N$) Yang-Mills theory and QCD, it is crucial to identify the reason for the technical complications in naive Kogut-Susskind-based approaches and provide alternative lattice formulations. In fact, it is not difficult to understand what makes the Kogut-Susskind Hamiltonian so complicated; a key observation is that theories based on non-compact variables, such as scalar quantum field theory~\cite{Jordan:2012xnu} and Hermitian matrix models~\cite{Gharibyan:2020bab}, can be simulated straightforwardly. This simple fact leads to a simple and profound guiding principle: \textit{use non-compact variables, and avoid compact variables}, because \textit{the complicated structure in the Kogut-Susskind Hamiltonian comes from the use of compact variables}. Once having this guiding principle, it is natural to use the orbifold lattice Hamiltonian~\cite{Buser:2020cvn,Bergner:2024qjl,Kaplan:2002wv}, which is formulated using the non-compact variables. Indeed, the orbifold lattice Hamiltonian leads to exponential improvement in the quantum simulations of Yang-Mills theory and QCD, for any SU($N$) and any spatial dimensions~\cite{Halimeh:2024bth,Hanada:2025yzx,Halimeh:2025ivn}.
For example, a quantum circuit for Hamiltonian time evolution via Trotter decomposition has been constructed explicitly, for arbitrary truncation level and using only analytic methods, from CNOT gates, one-qubit rotations and quantum Fourier transforms, with a number of gates that grows polynomially in $Q$ --- as $Q^{4}$ per site and per Trotter step for the orbifold lattice, see Sec.~6 of ref.~\cite{Halimeh:2024bth} --- and without relying on a black box~\cite{Halimeh:2024bth,Halimeh:2025ivn}. Such strong results are obtained because of a simple, universal structure of the Hamiltonian that is agnostic to the details of the theories such as gauge group or spatial dimensions.

In this work, we present a synthesis of the two approaches: the Kogut-Susskind Hamiltonian emerges as a limiting case from the orbifold lattice Hamiltonian. From this point of view, the orbifold lattice formulation appears as a controlled relaxation of the constraints of the field-theoretical target space. 
The equivalence of the Kogut-Susskind and the orbifold lattice Hamiltonians in the limit of infinite scalar mass is trivial in the continuum and at weak coupling. This paper aims to test a smooth transition at finite lattice spacing and strong couplings as well.  

That we can obtain the Kogut-Susskind Hamiltonian as a special limit of the orbifold lattice Hamiltonian means that the significant advantage of the latter ---  that \textit{one can easily write the truncated Hamiltonian explicitly by hand, and efficient quantum circuits can be designed by hand} --- can be transferred to the former. Therefore, we can enjoy the advantages of the standard Kogut-Susskind formulation, in particular related to theoretical considerations ---  there is a direct correspondence with Wilson's Euclidean path-integral formulation on the lattice~\cite{Wilson:1974sk}, which is a framework that has underpinned countless investigations --- resolving the technical disadvantages in its application to quantum simulations. 

This paper demonstrates that the Kogut-Susskind Hamiltonian can be derived as a specific limiting case of the orbifold lattice Hamiltonian, precisely when the mass parameters of certain fields approach infinity, without introducing additional lattice artifacts. Crucially, this limit can be systematically approached without compromising the quantum simulation advantages inherent to the orbifold lattice formulation. This realization offers a methodological synthesis: the quantum simulation of the Kogut-Susskind Hamiltonian becomes achievable through the orbifold lattice Hamiltonian evaluated at various mass parameters, with subsequent extrapolation to the infinite-mass limit. Given that the orbifold lattice Hamiltonian facilitates exponentially faster quantum simulations than conventional Kogut-Susskind approaches, our method delivers an exponential computational acceleration. The elegance of this approach lies in its universal applicability across arbitrary gauge groups and spacetime dimensions.\footnote{
Historically, ref.\cite{Kaplan:2002wv} introduced the orbifold lattice construction by performing an orbifold projection to the Banks-Fishler-Shenker-Susskind (BFSS) matrix model~\cite{Banks:1996vh} to build a lattice action of super Yang-Mills theory preserving a part of supersymmetry. The simplicity of the orbifold lattice Hamiltonian derives directly from the underlying simplicity of the BFSS Hamiltonian. Our proposal, therefore, represents a conceptual bridge that harnesses the simplicity of Banks-Fishler-Shenker-\underbar{Susskind} Hamiltonian to transcend the technical limitations inherent in the Kogut-\underbar{Susskind} Hamiltonian.
}

While this equivalence has been previously suggested~\cite{Buser:2020cvn,Bergner:2024qjl}, comprehensive details were not provided, as the focus was primarily on establishing the equivalence in the continuum limit through radiative corrections rather than demonstrating the equivalence at finite lattice spacing and outside the weak coupling regime. This paper provides a rigorous analysis of the lattice-regularized equivalence. Following a general analytical treatment for SU($N$) Yang-Mills theory applicable to arbitrary $N$ and dimensions with straightforward extensions to QCD, we present numerical evidence for the $(2+1)$-dimensional pure Yang-Mills theory. Our computational approach employs the path-integral formulation, which, while equivalent to the Hamiltonian formulation, offers complementary computational advantages. For quantitative investigation, we employ lattice Monte Carlo methods to compare $(2+1)$-d SU($N$) Wilson action with $(2+1)$-d SU($N$) orbifold action under identical lattice spacing and volume conditions for $N=2$ and $3$. Here, we adopt the Euclidean signature, which is convenient for our purpose of seeing how heavy degrees of freedom affect physics at low energy. Within the orbifold-lattice framework, we examine multiple bare scalar mass values, confirming that Wilson action results emerge naturally through extrapolation to infinite bare mass. This methodology enables a quantitative assessment of convergence to the Kogut-Susskind Hamiltonian.

Note, however, that \textit{there is no reason that we have to take the Kogut-Susskind limit}; the orbifold-lattice formulation leads to the same continuum limit as the Kogut-Susskind formulation, because the scalars acquire positive radiative mass corrections (generic for scalars coupled to gauge fields at constant bare mass) that grow relative to the dynamical scale as $a\to 0$, so the scalars become heavy and decouple from the low-energy dynamics. What we demonstrate in this paper is that the Kogut-Susskind formulation can be reproduced from the orbifold lattice including unphysical lattice artifacts, if one prefers to do that; see Fig.~\ref{fig:continuum-and KS-limits}. Note also that no new energy scale beyond the cutoff scale has to be introduced to estimate the Kogut-Susskind limit. Still, there may be a nontrivial dynamics that causes complications for quantum simulations, for example the use of smaller Trotter steps; but that is also a potential problem of the original Kogut-Susskind approach. 

\begin{figure}[htbp]
\begin{center}
\begin{tikzpicture}[node distance=2cm]
% nodes
\node (OL) [theory] {Orbifold lattice $\hat{H}$};
\node (KS) [theory, right=5cm of OL] {Kogut-Susskind};
\node (YM) [theory, below=of KS, xshift=-4.2cm] {Yang-Mills};
% Arrows
\draw [arrow] (OL) -- (KS);
\draw [arrow] (OL) -- (YM);
\draw [arrow] (KS) -- (YM);
% Optional labels for arrows
\node [text width=4cm, align=center, above] at ($(OL)!0.51!(KS)$) {Kogut-Susskind limit};
\node [text width=4cm, align=center, below] at ($(OL)!0.51!(KS)$) {(optional)};
\node [whitelabel, text width=4cm, align=center] at ($(OL)!0.5!(YM)$) {continuum limit};
\node [whitelabel, text width=4cm, align=center] at ($(KS)!0.5!(YM)$) {continuum limit};
% Draw a surrounding box around the entire flowchart
\draw [rounded corners=5pt, thick] 
  ($(OL.north west)+(-0.6,0.8)$) rectangle 
  ($(KS.north east -| YM.south east)+(5,-3.8)$);
\end{tikzpicture}
\end{center}\caption{
Both orbifold lattice Hamiltonian and Kogut-Susskind Hamiltonian describe Yang-Mills theory in the continuum limit. Optionally, the Kogut-Susskind Hamiltonian can be obtained as a special limit of the orbifold lattice Hamiltonian --- which is different from the continuum limit --- including unphysical lattice artifacts.
}\label{fig:continuum-and KS-limits}
\end{figure}

The paper proceeds as follows: Section~\ref{sec:orbifold_Hamiltonian} introduces the orbifold lattice Hamiltonian for SU($N$) Yang-Mills theory. Section~\ref{sec:lattice_actions} presents relevant lattice actions and elucidates the relationship between orbifold lattice Hamiltonian/action and Kogut-Susskind Hamiltonian/Wilson action, with particular emphasis on the infinite-mass parameter limit. Section~\ref{sec:numerical_demonstration} provides numerical confirmations for $(2+1)$-d SU(2) and SU(3) theories. Finally, Section~\ref{sec:conclusion} synthesizes our findings, comments on the extension to QCD, and explores promising future research directions.
%%%%%%%%%%%%%%%%%%%%%%%%%%%
%%%%%%%%%%%%%%%%%%%%%%%%%%%
\section{Orbifold lattice Hamiltonian for SU($N$) Yang-Mills theory}\label{sec:orbifold_Hamiltonian}
%%%%%%%%%%%%%%%%%%%%%%%%%%%
%%%%%%%%%%%%%%%%%%%%%%%%%%%
In this section, we introduce the orbifold Hamiltonian for an SU($N$) Yang-Mills theory in $d+1$ dimensions ($d$ discrete spatial dimensions and continuous time). The spatial link variables are $N\times N$ complex matrices $Z_{j,\vec{n}}$, where $j=1,\cdots,d$ and $\vec{n}$ labels spatial points. 

The complex link variable $Z_{j,\vec{n}}$ can be decomposed into a positive-definite Hermitian matrix $W_{j,\vec{n}}$ and unitary link variable $U_{j,\vec{n}}$~\cite{Unsal:2005yh}:
\begin{align}
Z_{j,\vec{n}}=\sqrt{\frac{a^{d-2}}{2g_d^2}}W_{j,\vec{n}}U_{j,\vec{n}}\, . 
\label{eq:Z-W-U}
\end{align}
The unitary variable $U_{j,\vec{n}}$ describes the gauge field $A_j$, with the well-known relation $U_{j,\vec{n}}=\exp(\mathrm{i}ag_dA_{j,\vec{n}})$. Note that a priori the determinant of $U_{j,\vec{n}}$ is not fixed to one. Instead, the Hamiltonian has an additional term that forces the U(1) part $\mathrm{Tr}A_{j,\vec{n}}$ to be heavy and the determinant close to 1. The SU($N$) part will be treated as the gauge field, but not the U(1) part.\footnote{
In the past, mainly U($N$) was considered. In this paper, we consider SU($N$). See, e.g., ref.~\cite{Bergner:2024qjl}, that pointed out either SU($N$) and U($N$) can be gauged and then studied the U($N$) case.}$^,$\footnote{When the orbifold lattice was introduced in ref.~\cite{Kaplan:2002wv}, the motivation was to obtain lattice regularization of supersymmetric gauge theories keeping a part of supercharges intact. For that purpose, it was necessary to use U($N$). 
}  
The Hamiltonian will also have a term that forces $W_{j,\vec{n}}$ to be close to identity. $W_{j,\vec{n}}$ is related to 
a scalar field $\phi_j$ by $W_{j,\vec{n}}=\exp\left(ag_d\phi_{j,\vec{n}}\right)$. 

Let $\bar{Z}_{j,\vec{n}}$ be the Hermitian conjugate of $N\times N$ matrix $Z_{j,\vec{n}}$, i.e., $(\bar{Z}_{j,\vec{n}})_{ab}=[(Z_{j,\vec{n}})_{ba}]^\ast$. 
We take $P_{j,\vec{n}}$ and $\bar{P}_{j,\vec{n}}$ to be the conjugate momenta of $\bar{Z}_{j,\vec{n}}$ and $Z_{j,\vec{n}}$. 
For these operators, the commutation relations are 
\begin{align}
[\hat{Z}_{j,\vec{n},pq},\hat{\bar{P}}_{k\vec{n}',rs}]=\mathrm{i}\delta_{jk}\delta_{\vec{n}\vec{n}'}\delta_{ps}\delta_{qr},
\label{eq:commutation-relation-orbifold}
\end{align}
and
\begin{align}
[\hat{Z},\hat{P}]
=
[\hat{\bar{Z}},\hat{\bar{P}}]
=
[\hat{Z},\hat{Z}]
=
[\hat{Z},\hat{\bar{Z}}]
=
[\hat{\bar{Z}},\hat{\bar{Z}}]
=
[\hat{P},\hat{P}]
=
[\hat{P},\hat{\bar{P}}]
=
[\hat{\bar{P}},\hat{\bar{P}}]
=
0\, . 
\end{align}
The Hamiltonian is 
\begin{align}
\hat{H}
&=
\sum_{\vec{n}}
{\rm Tr}\Biggl(
\sum_{j=1}^d \hat{P}_{j,\vec{n}} \,  \hat{\bar{P}}_{j,\vec{n}}
+
%% second term
\frac{g_d^2}{2a^d}\left|\sum_{j=1}^d
\left(
\hat{Z}_{j,\vec{n}} \, \hat{\bar{Z}}_{j,\vec{n}} -\hat{\bar{Z}}_{j,\vec{n}-\hat{j}}\hat{Z}_{j,\vec{n}-\hat{j}}
\right)
\right|^2 
\nonumber\\
&\qquad\qquad\qquad
%%Second line
%%
+
\frac{2g_d^2}{a^d}\sum_{j<k}
\left|
\hat{Z}_{j,\vec{n}} \, \hat{Z}_{k,\vec{n}+\hat{j}}
-
\hat{Z}_{k,\vec{n}} \, \hat{Z}_{j,\vec{n}+\hat{k}}
\right|^2
 \Biggl)
 \, +\,  
 \Delta\hat{H}\, , 
\label{eq:Hamiltonian_orbifold}
\end{align}
where 
\begin{align}
\Delta\hat{H}
=
\hat{H}_{\rm mass}
+
\hat{H}_{\rm linear}\, , 
\end{align}
%%%
\begin{align}
\hat{H}_{\rm mass}
&=
\frac{m^2g_d^2}{2a^{d-2}}\sum_{\vec{n}}\sum_{j=1}^d
{\rm Tr}
\left|\hat{Z}_{j,\vec{n}}\hat{\bar{Z}}_{j,\vec{n}} -\frac{a^{d-2}}{2g_d^2}\right|^2
\nonumber\\
&\quad
+
\frac{m^2_{\rm U(1)}a^{d-2}}{2g_d^2}
\sum_{\vec{n}}\sum_{j=1}^d
\left|
    \left(\frac{a^{d-2}}{2g_d^2}\right)^{-N/2}\det(\hat{Z}_{j,\vec{n}})-1
    \right|^2\, ,  
    \label{eq:H_add}
\end{align}
and
\begin{align}
\hat{H}_{\rm linear}
=
-\gamma\sum_{\vec{n}}\sum_{j=1}^d{\rm Tr}\left(\hat{Z}_{j,\vec{n}}\hat{\bar{Z}}_{j,\vec{n}}\right)\, . 
    \label{eq:H_add_linear}
\end{align}

The meaning of the different contributions of the Hamiltonian are explained in ref.~\cite{Bergner:2024qjl}. In this reference, we have included a term $\sum_{\vec{n}}\sum_{j=1}^d
\left| \frac{1}{N}{\rm Tr}(\hat{Z}_{j,\vec{n}}\hat{\bar{Z}}_{j,\vec{n}}) -\frac{a^{d-2}}{2g_d^2}\right|^2$ which would add a mass to the U(1) part of $\phi_j$, but not to the U(1) part of $A_j$. We have replaced this term by the second term in \eqref{eq:H_add}, which corresponds to a mass for both U(1) parts.

Using $^{\rm (R)}$ and $^{\rm (I)}$ to denote real and imaginary parts\footnote{
Alternatively, we could take $^{\rm (R)}$ and $^{\rm (I)}$ to be Hermitian and anti-Hermitian parts. 
} of $\hat{Z}$ as $\hat{Z}=\frac{\hat{Z}^{\rm (R)}+\mathrm{i}\hat{Z}^{\rm (I)}}{\sqrt{2}}$,
we obtain 
\begin{align}
[\hat{Z}^{\rm (R)}_{j,\vec{n},pq},\hat{P}^{\rm (R)}_{k,\vec{n}',rs}]
=
[\hat{Z}^{\rm (I)}_{j,\vec{n},pq},\hat{P}^{\rm (I)}_{k,\vec{n}',rs}]
=
\mathrm{i}\delta_{jk}\delta_{\vec{n}\vec{n}'}\delta_{ps}\delta_{qr}\, . 
\end{align}

The Hilbert space is defined by using the coordinate eigenstates $\ket{Z}$:
\begin{align}
\mathcal{H}_{\rm ext}
=
\mathrm{Span}\left\{
\ket{Z}\ ; \hat{Z}_{j,\vec{n}}\ket{Z}=Z_{j,\vec{n}}\ket{Z}
\right\}\, .
\end{align}
Specifically, we consider the states $\ket{\Phi}=\int dZ\Phi(Z)\ket{Z}$ with the square-integrable wave function $\Phi(Z)$. Note that $\mathcal{H}_{\rm ext}$ is the extended Hilbert space that contains the gauge non-singlet states. A symmetry of the Hamiltonian $\hat{H}$ can be gauged if we take only singlet states or avoid double-counting of the states related by the symmetry. The Hamiltonian is invariant under U($N$) transformation, but only the SU($N$) subgroup is gauged.

Following \eqref{eq:Z-W-U}, the complex link variable $Z_{j,\vec{n}}$ is decomposed into $W_{j,\vec{n}}$ and $U_{j,\vec{n}}$. 
Since 
\begin{align}
Z_{j,\vec{n}}\bar{Z}_{j,\vec{n}}=\frac{a^{d-2}}{2g_d^2}W^2_{j,\vec{n}}\, , 
\end{align}
the first term of $\hat{H}_{\rm mass}$ in \eqref{eq:H_add} pushes $W_{j,\vec{n}}$ close to the identity. In the continuum limit, at tree level, we obtain a mass term of scalar $\phi_j$ proportional to $\sum_j\mathrm{Tr}\phi_j^2$. 
The second term in \eqref{eq:H_add} leads to the mass of the U(1) part,  
$\sum_{j=1}^d
\left(
({\rm Tr}\phi_j)^2
+
({\rm Tr}A_j)^2
\right)$. 

When the vacuum expectation value of the scalar field is nonzero, the lattice spacing shifts effectively from $a$; see e.g.,~ref~\cite{Bergner:2026duh}. An optional term $\hat{H}_{\rm linear}$ defined by \eqref{eq:H_add_linear} adds a linear term $-\gamma\sum_j\mathrm{Tr}\phi_j$ that can be used to set the expectation value to be zero and the `bare' lattice spacing $a$ agrees precisely with the effective lattice spacing. This tuning of the effective spatial lattice spacing can be chosen such that it leads to an improved convergence towards the isotropic Kogut-Susskind limit. Therefore $\hat{H}_{\rm linear}$ can be considered as an improvement term.
The rest of the Hamiltonian describes Yang-Mills theory coupled to scalars.

When $m^2$ and $m^2_{\rm U(1)}$ are large (and, optionally, when $\gamma$ is tuned to an appropriate value), $\det U_{j,\vec{n}}$ and $W_{j,\vec{n}}$ are well localized around 1 and the identity matrix $\textbf{1}_N$, respectively. The scalars $\phi_j$ and the U(1) part of $A_j$ decouple,\footnote{
More precisely, they behave as harmonic oscillators with parametrically large frequency, which get stuck in the ground state.
} and only the SU($N$) gauge field is left, leading to the Kogut-Susskind Hamiltonian for SU($N$) pure Yang-Mills theory. In Section~\ref{sec:equivalence_analytic}, we will study this limit quantitatively by using the Euclidean path integral.\footnote{Note that $\left|\sum_{j=1}^d
\left(
\hat{Z}_{j,\vec{n}} \, \hat{\bar{Z}}_{j,\vec{n}} -\hat{\bar{Z}}_{j,\vec{n}-\hat{j}}\hat{Z}_{j,\vec{n}-\hat{j}}
\right)
\right|^2 $ vanishes in this limit. Therefore, it might be better to omit this term for quantum simulations in this limit.
}

The orbifold lattice Hamiltonian \eqref{eq:Hamiltonian_orbifold} belongs to a class of Hamiltonians of bosonic systems of the form
\begin{align}
    \hat{H}
    =
    \sum_a \frac{\hat{p}_a^2}{2}
    +
    V(\hat{x}_1,\hat{x}_2,\cdots)\, , 
    \label{generic_class}
\end{align}
where $\hat{x}_a$ and $\hat{p}_a$ are the coordinate and momentum operators of the $a$-th boson that satisfy the canonical commutation relations
\begin{align}
    [\hat{x}_a,\hat{p}_b]
    =
    \mathrm{i}\delta_{ab}\, . 
\end{align}
We assume the potential part $V$ not to be complicated, e.g., a polynomial or analytic function that can be well approximated by a lower-order truncated Taylor series. In the case of the orbifold lattice Hamiltonian studied in this paper, $V$ is a polynomial of degree $2N$. This class of theories is simple enough to admit efficient quantum simulation algorithms~\cite{Halimeh:2024bth}.

%%%%%%%%%%%%%%%%%%%%%%%%%%%
%%%%%%%%%%%%%%%%%%%%%%%%%%%
\section{Kogut-Susskind Hamiltonian from the orbifold lattice}\label{sec:lattice_actions}
%%%%%%%%%%%%%%%%%%%%%%%%%%%
%%%%%%%%%%%%%%%%%%%%%%%%%%%
We can see the relationship between the Kogut-Susskind Hamiltonian and the infinite-mass limit of the orbifold lattice in the following way. Let us focus on the scalars, because the same argument applies to the U(1) part. When the mass is large, it is reasonable to assume that the scalars are approximated well by keeping only the free part (momentum term and the mass term); indeed, we can justify this assumption shortly. If the mass is larger than the energy scale under consideration, then the scalars are frozen to the ground state of the harmonic oscillators. 
In the coordinate basis, using the analytic formula for the ground state, we can see that the width of the ground-state wave function, $\sqrt{\langle\phi^{2}\rangle}=(2m)^{-1/2}$, decays as $m^{-1/2}$. Because of this suppression, the interaction can be neglected, indeed. Therefore, we conclude that the theory reduces to Kogut-Susskind plus decoupled harmonic oscillators.

An easy way to see the connection between the Kogut-Susskind Hamiltonian and the orbifold lattice Hamiltonian quantitatively, even at finite mass, is to switch to the path-integral formalism. Because we want to see how heavy degrees of freedom affect physics at low energy, it is convenient to consider the Euclidean signature.
With the Euclidean signature, we can introduce a spacetime lattice so that the path integral reduces to a usual integral with a finite number of variables, which brings the proof of the equivalence into a form accessible by numerical methods. We can use Monte Carlo simulations to obtain quantitative results. 

In this section, we provide the two spacetime lattice actions to demonstrate the emergence of the Kogut-Susskind Hamiltonian from the orbifold lattice Hamiltonian. The first is the standard Wilson action~\cite{Wilson:1974sk} (Section~\ref{sec:Wilson_action}) that corresponds to the Kogut-Susskind Hamiltonian when the continuum limit is taken along the time direction. The second is the orbifold lattice action~\cite{Bergner:2024qjl} (Section~\ref{sec:orbifold_action}). This is slightly different from those in the original papers~\cite{Cohen:2003xe,Cohen:2003qw,Kaplan:2005ta} reflecting a difference of motivation (the target of the original papers was exact supersymmetry on the lattice). Specifically, we use the complex link variables only for spatial links, and, as a consequence, we can take the gauge group of these links to be SU($N$) rather than U($N$). As already explained, the spatial links are complex matrices, which get close to SU($N$) once the masses get sufficiently large. 

In Section~\ref{sec:equivalence_analytic}, we show the equivalence of the two theories directly at the regularized level, without taking the continuum limit in either temporal or spatial dimensions. This equivalence guarantees the equivalence of the Hamiltonian formulations when the continuum limit is taken along the time direction. 
%%%%%%%%%%%%%%%%%%%%%%%%%%%
%%%%%%%%%%%%%%%%%%%%%%%%%%%
\subsection{Wilson action}\label{sec:Wilson_action}
%%%%%%%%%%%%%%%%%%%%%%%%%%%
%%%%%%%%%%%%%%%%%%%%%%%%%%%
From this section on, $\vec{n}$ labels the sites of the $(d+1)$-dimensional spacetime lattice, and sums over $\vec{n}$ include the Euclidean time direction.
The Wilson action for SU($N$) Yang-Mills theory in $(d+1)$-dimensional theory is given by 
\begin{align}
S_{\rm Wilson}
&=
\sum_{\vec{n}}
{\rm Tr}\Biggl(
-\frac{1}{a_t}
\frac{a^{d-2}}{2g_d^2}
\sum_{j=1}^d
\left(
U_{t,\vec{n}}U_{j,\vec{n}+\hat{t}}U_{t,\vec{n}+\hat{j}}^\dagger U^\dagger_{j,\vec{n}}
+
U_{t,\vec{n}}^\dagger U_{j,\vec{n}}U_{t,\vec{n}+\hat{j}}
U^\dagger_{j,\vec{n}+\hat{t}}
\right)
\nonumber\\
&\qquad\qquad
-
\frac{a_ta^{d-4}}{2g_d^2}
\sum_{j<k}
\left(
U_{j,\vec{n}}U_{k,\vec{n}+\hat{j}}
U^\dagger_{j,\vec{n}+\hat{k}}
U^\dagger_{k,\vec{n}}
+
U_{k,\vec{n}}U_{j,\vec{n}+\hat{k}}
U^\dagger_{k,\vec{n}+\hat{j}}
U^\dagger_{j,\vec{n}}
\right)
 \Biggl)\, .  
\end{align}
Here, $a_t$ and $a$ are temporal and spatial lattice spacings, respectively. We introduced different spacings (anisotropic lattice) such that we can take a limit $a_t\to 0$ that corresponds to the Kogut-Susskind Hamiltonian. 
As a path-integral measure, we use the Haar measure for both $U_t$ and $U_j$. 
%%%%%%%%%%%%%%%%%%%%%%%%%%%
%%%%%%%%%%%%%%%%%%%%%%%%%%%
\subsection{Orbifold lattice action}\label{sec:orbifold_action}
%%%%%%%%%%%%%%%%%%%%%%%%%%%
%%%%%%%%%%%%%%%%%%%%%%%%%%%
The orbifold lattice action for SU($N$) Yang-Mills theory in $d+1$ dimensions that follows from~\eqref{eq:Hamiltonian_orbifold}, see also \cite{Bergner:2024qjl}, is given as
\begin{align}
S_{\rm orbifold}
&=
\sum_{\vec{n}}
{\rm Tr}\Biggl(
  \frac{1}{a_t}\sum_{j=1}^d
\left|
U_{t,\vec{n}}Z_{j,\vec{n}+\hat{t}}
-
Z_{j,\vec{n}}U_{t,\vec{n}+\hat{j}}
\right|^2
\nonumber\\
&\qquad\qquad
+
\frac{g_d^2a_t}{2a^d}\left|\sum_{j=1}^d
\left(
Z_{j,\vec{n}} \bar{Z}_{j,\vec{n}} -\bar{Z}_{j,\vec{n}-\hat{j}}Z_{j,\vec{n}-\hat{j}}
\right)
\right|^2 
\nonumber\\
&\qquad\qquad
+
\frac{2g_d^2a_t}{a^d}\sum_{j<k}
\left|
Z_{j,\vec{n}}Z_{k,\vec{n}+\hat{j}}
-
Z_{k,\vec{n}}Z_{j,\vec{n}+\hat{k}}
\right|^2
 \Biggl)
 \ +\ \Delta S_{\rm orbifold}\, , 
\label{eq:4d_lattice-action}
\end{align}
\begin{align}
\Delta S_{\rm orbifold}
&\equiv
\frac{m^2g_d^2a_ta^{2-d}}{2}\sum_{\vec{n}}\sum_{j=1}^d
{\rm Tr}
\left| Z_{j,\vec{n}}\bar{Z}_{j,\vec{n}} -\frac{a^{d-2}}{2g_d^2}\right|^2
\nonumber\\
&\quad
+
\frac{m^2_{\rm U(1)}a_ta^{d-2}}{2g_d^2}
\sum_{\vec{n}}\sum_{j=1}^d
\left|
    \left(\frac{a^{d-2}}{2g_d^2}\right)^{-N/2}\det(Z_{j,\vec{n}})-1
\right|^2
\nonumber\\
&\quad
-
a_t\gamma\sum_{\vec{n}}\sum_{j=1}^d{\rm Tr}\left(Z_{j,\vec{n}}\bar{Z}_{j,\vec{n}}\right)\, .
\label{eq:deltaS_orbifold}
\end{align}
%Optionally, we can add a term proportional to $\sum_{\vec{n}}\sum_{j=1}^3\mathrm{Tr}(Z_{j,\vec{n}} \bar{Z}_{j,\vec{n}})$, but we do not consider this term in this paper. 

We take the unitary temporal link variable $U_{{\rm t},\vec{n}}$ as elements of the gauge group SU($N$) and \textit{not} U($N$). 
The complex link variable $Z_{j,\vec{n}}$ can be decomposed as in \eqref{eq:Z-W-U}. 
The additional term $\Delta S_{\rm orbifold}$ forces $W_{j,\vec{n}}$ and $\det U_{j,\vec{n}}$ to fluctuate around the identity. The tree level continuum limit $a\to 0$ of the action in terms of adjoint scalar $\phi_j$ and gauge field $A_j$ is derived using $W_{j,\vec{n}}=\exp\left(ag_d\phi_{j,\vec{n}}\right)$ and $U_{j,\vec{n}}=\exp\left(\mathrm{i}ag_dA_{j,\vec{n}}\right)$.
This assumes that $W_{j,\vec{n}}$ fluctuates around the identity matrix $\textbf{1}_N$; a small departure from the identity can be absorbed into a redefinition of lattice spacing, as demonstrated in detail in ref.~\cite{Bergner:2026duh}. Optionally, the last term in \eqref{eq:deltaS_orbifold} can be used to eliminate such a departure by tuning $\gamma$, without taking $m^2$ large~\cite{Halimeh:2024bth,Bergner:2026duh}. Either way, the corresponding continuum action at tree level is
\begin{align}
S_{\rm orbifold}
=
\int \mathrm{d}^{d+1}x{\rm Tr}
\left(
\frac{1}{4}F_{\mu\nu}^2
+
\frac{1}{2}(D_\mu \phi_I)^2
+
\frac{g_d^2}{4}[\phi_I,\phi_J]^2
\right)
\end{align}

\begin{align}
\Delta S_{\rm orbifold}
&=
\frac{m^2}{2}\int\mathrm{d}^{d+1}x
\sum_{j=1}^d\, {\rm Tr}(\phi_j^2)
\nonumber\\
&\quad
+
\frac{m^2_{\rm U(1)}}{2}\int\mathrm{d}^{d+1}x
\sum_{j=1}^d
\left(
({\rm Tr}\phi_j)^2
+
({\rm Tr}A_j)^2
\right)\, . 
\end{align}
Note that, as the path-integral measure, we use the Haar measure for $U_t$ and the flat measure for $Z_j$. 
%%%%%%%%%%%%%%%%%%%%%%%%%%%
%%%%%%%%%%%%%%%%%%%%%%%%%%%
\subsection{Relationship between the two theories}\label{sec:equivalence_analytic}
%%%%%%%%%%%%%%%%%%%%%%%%%%%
%%%%%%%%%%%%%%%%%%%%%%%%%%%
To remove the scalars $\phi_j$ and the U(1) part of $A_j$, we send $m^2$ and $m^2_{\rm U(1)}$ to infinity. In this limit,\footnote{
When $W$ fluctuates about $c\cdot\textbf{1}$ with $c\neq 1$, lattice spacings $a$ and $a_t$ shift effectively; see ref.~\cite{Bergner:2026duh}.
} 
$W\to\mathbf{1}_N$,   $Z\to\sqrt{\frac{a^{d-2}}{2g_d^2}}U$, and $\det U\to 1$. The orbifold lattice action \eqref{eq:4d_lattice-action} becomes 

\begin{align}
S_{\rm orbifold}
&=
\sum_{\vec{n}}
{\rm Tr}\Biggl(
\frac{1}{a_t}
\frac{a^{d-2}}{2g_d^2}
\sum_{j=1}^d
\left|
U_{t,\vec{n}}U_{j,\vec{n}+\hat{t}}
-
U_{j,\vec{n}}U_{t,\vec{n}+\hat{j}}
\right|^2
\nonumber\\
&\qquad\qquad
+
\frac{g_d^2a_t}{2a^d}
\left(
\frac{a^{d-2}}{2g_d^2}
\right)^2
\left|
\sum_{j=1}^d
\left(
U_{j,\vec{n}}U^\dagger_{j,\vec{n}} -U^\dagger_{j,\vec{n}-\hat{j}}U_{j,\vec{n}-\hat{j}}
\right)
\right|^2 
\nonumber\\
&\qquad\qquad
+
\frac{2g_d^2a_t}{a^d}
\left(
\frac{a^{d-2}}{2g_d^2}
\right)^2
\sum_{j<k}
\left|
U_{j,\vec{n}}U_{k,\vec{n}+\hat{j}}
-
U_{k,\vec{n}}U_{j,\vec{n}+\hat{k}}
\right|^2
 \Biggl)\, .  
\end{align}
The second line vanishes since $U_{j,\vec{n}}U^\dagger_{j,\vec{n}} = U^\dagger_{j,\vec{n}-\hat{j}}U_{j,\vec{n}-\hat{j}}=\mathbf{1}_N$. 
The third line is
\begin{align}
\lefteqn{
\frac{a_ta^{d-4}}{2g_d^2}
\sum_{j<k}
{\rm Tr}\left|
U_{j,\vec{n}}U_{k,\vec{n}+\hat{j}}
-
U_{k,\vec{n}}U_{j,\vec{n}+\hat{k}}
\right|^2
}\nonumber\\
 &=
\frac{a_ta^{d-4}}{2g_d^2}
\sum_{j<k}
{\rm Tr}\left(
2\cdot\mathbf{1}_N
-
U_{j,\vec{n}}U_{k,\vec{n}+\hat{j}}
U^\dagger_{j,\vec{n}+\hat{k}}
U^\dagger_{k,\vec{n}}
-
U_{k,\vec{n}}U_{j,\vec{n}+\hat{k}}
U^\dagger_{k,\vec{n}+\hat{j}}
U^\dagger_{j,\vec{n}}
\right)\, ,  
\end{align}
which is the standard plaquette terms.
Likewise, the first line is also written in terms of plaquette. 
Therefore, up to an additive constant, 
\begin{align}
S_{\rm orbifold}
&=
\sum_{\vec{n}}
{\rm Tr}\Biggl(
-\frac{1}{a_t}
\frac{a^{d-2}}{2g_d^2}
\sum_{j=1}^d
\left(
U_{t,\vec{n}}U_{j,\vec{n}+\hat{t}}U_{t,\vec{n}+\hat{j}}^\dagger U^\dagger_{j,\vec{n}}
+
U_{t,\vec{n}}^\dagger U_{j,\vec{n}}U_{t,\vec{n}+\hat{j}}
U^\dagger_{j,\vec{n}+\hat{t}}
\right)
\nonumber\\
&\qquad\qquad
-
\frac{a_ta^{d-4}}{2g_d^2}
\sum_{j<k}
\left(
U_{j,\vec{n}}U_{k,\vec{n}+\hat{j}}
U^\dagger_{j,\vec{n}+\hat{k}}
U^\dagger_{k,\vec{n}}
+
U_{k,\vec{n}}U_{j,\vec{n}+\hat{k}}
U^\dagger_{k,\vec{n}+\hat{j}}
U^\dagger_{j,\vec{n}}
\right)
 \Biggl)\, .  
\end{align}
This is the same as the Wilson action. The measure of the integration of the unitary part arising from the flat measure is the Haar measure, which is the same as the measure used for the path integral with the Wilson action. Therefore, we obtain exactly the same path-integral weight and path-integral measure from the Wilson action and from the infinite-mass limit of the orbifold lattice action.

To see if such a limit is practically under control, numerical Monte Carlo simulations are required. In the next section, we study SU(2) and SU(3) pure Yang-Mills theory in $2+1$ dimensions and confirm that it is straightforward to take this limit.  
%%%%%%%%%%%%%%%%%%%%%%%%%%%
%%%%%%%%%%%%%%%%%%%%%%%%%%%
\section{Lattice simulations for $(2+1)$-d SU(2) and SU(3) theory}\label{sec:numerical_demonstration}
%%%%%%%%%%%%%%%%%%%%%%%%%%%
%%%%%%%%%%%%%%%%%%%%%%%%%%%
We have seen that the SU($N$) Wilson action and the Haar measure of the path integral are obtained from the orbifold lattice action and flat measure by sending $m^2$ and $m^2_{\rm U(1)}$ to infinity. In practice, for quantum simulations, we should take several values of $m^2$ and $m^2_{\rm U(1)}$ and then extrapolate the results to the infinite mass limit. 

In this section, we demonstrate such an extrapolation for $(2+1)$-d Yang-Mills theory, with gauge group SU(2) and SU(3). Specifically, we use the Hybrid Monte Carlo algorithm~\cite{Duane:1987de} (see ref.~\cite{Hanada_Matsuura_2022} for an introductory review) for the demonstration.\footnote{
Simulation codes are available at \url{https://github.com/masanorihanada/3d_pure_YM_Wilson_action} and \url{https://github.com/masanorihanada/3d_orbifold_lattice_YM}. 
}
We want to provide evidence for the equivalence even at the regularized level, considering small lattice sizes without the continuum extrapolation.

Note that the Yang-Mills coupling constant is dimensionful at $d\neq 3$, i.e., $g^2_d$ has mass dimension $3-d$, and the lattice spacing should be measured in units of the coupling constant. To take the $N$ dependence into account, the 't Hooft coupling $g^2_dN$ provides a typical energy scale. Dimensionless combinations are  $(g^2_dN) a^{3-d}$ and $(g^2_dN) a_t^{3-d}$. If these dimensionless combinations are small, the system is close to the continuum limit. 
We set the coupling constant to $g^2_{d=2}=1$. Therefore, dimensionful quantities like lattice spacing $a$ or temperature $T$ should be interpreted as the dimensionless combinations $ag^2_{d=2}$ or $T/g^2_{d=2}$. Furthermore, we take $m^2=m_{\rm U(1)}^2$ and study the limit of infinite mass. Whether these parameters are `large' or `small' should be considered in relation to the mass scale set by the coupling constant.

We will study two cases: with or without linear counter term $\hat{H}_{\rm linear}$. 
%%%%%%%%%%%%%%%%%%%%%%%%%%%
%%%%%%%%%%%%%%%%%%%%%%%%%%%
\subsection{Emergence of Kogut-Susskind as heavy mass limit of the orbifold theory}
%%%%%%%%%%%%%%%%%%%%%%%%%%%
%%%%%%%%%%%%%%%%%%%%%%%%%%%
As a first step, we study the Kogut-Susskind (KS) limit without introducing the improvement term $\hat{H}_{\rm linear}$, i.~e., setting $\gamma$ zero. 
The aim of this investigation is to demonstrate that the Kogut-Susskind results are reproduced even in the ultraviolet regime including visible lattice artefacts. For simplicity, we consider the isotropic KS action at zero temperature. A more precise comparison in the spirit of the original orbifold approach would include a dynamically determined anisotropy, which we postpone to a later publication. At finite temperature we consider an anisotropic case, but only as an alternative method to adjust the temperature.  We will use the following quantities for the confirmation of convergence to the KS limit:
\begin{itemize}
    \item 
To confirm agreement in the ultraviolet regime, we compute the spatial plaquettes. We use the one made of complex links, 
$\mathrm{Tr}\left(Z_{1,\vec{n}}Z_{2,\vec{n}+\hat{1}}\bar{Z}_{1,\vec{n}+\hat{2}}\bar{Z}_{2,\vec{n}}\right)$, and the one made of unitary links, 
$\mathrm{Tr}\left(U_{1,\vec{n}}U_{2,\vec{n}+\hat{1}}U^\dagger_{1,\vec{n}+\hat{2}}U^\dagger_{2,\vec{n}}\right)$. 
Note that $U_{j,\vec{n}}$ can be obtained from $Z_{j,\vec{n}}$ using \eqref{eq:Z-W-U}. 
As $m^2=m_{\rm U(1)}^2\to\infty$, these plaquette values should converge to the spatial plaquette $\mathrm{Tr}\left(U_{1,\vec{n}}U_{2,\vec{n}+\hat{1}}U^\dagger_{1,\vec{n}+\hat{2}}U^\dagger_{2,\vec{n}}\right)$ obtained with the Wilson action at the same parameters up to an overall factor (4 for the former and 1 for the latter, specific to the $2+1$-dimensional case studied here; in higher dimensions this factor is $a$-dependent).
\footnote{As usual, the expectation values are obtained by averaging over spacetime points $\vec{n}$ and Monte-Carlo samples. They are denoted by $\langle\mathrm{Tr}(ZZ\bar{Z}\bar{Z})\rangle$ and $\langle\mathrm{Tr}(UUU^\dagger U^\dagger)\rangle_{\rm spatial}$, respectively. The temporal plaquette is, in addition, averaged over the two spacial directions $j=1,2$.}
    \item 
The temporal plaquette $\mathrm{Tr}\left(U_{t,\vec{n}}U_{j,\vec{n}+\hat{t}}U^\dagger_{t,\vec{n}+\hat{j}}U^\dagger_{j,\vec{n}}\right)$ averaged over $j=1,2$. The corresponding average value is denoted as  $\langle\mathrm{Tr}(UUU^\dagger U^\dagger)\rangle_{\rm temporal}$. The comparison of spatial and temporal plaquettes indicates the residual anisotropy.
    \item  Another important observable is the Polyakov loop. It can be used to identify the signal for the deconfinement transition. 
    Specifically, we consider the bare Polyakov loop without renormalization. Although it depends on the details of the ultraviolet regime through the renormalization factor, the general phase diagram detected by this quantity does not. Therefore, agreement of the bare Polyakov loop provides strong evidence for the agreement with the Kogut-Susskind action in the ultraviolet and infrared. 
    \item 
We also compute $\mathrm{Tr}\left(W_{j,\vec{n}}-\textbf{1}_N\right)^2$ and $\det U_{j,\vec{n}}$. We take the average over $j=1,2$, spacetime points $\vec{n}$, and samples with the corresponding averages $\langle\mathrm{Tr}(W-\textbf{1}_N)^2\rangle$ and $\langle\det U\rangle$. These averages should converge to 0 and 1, respectively in the KS limit.
\end{itemize}

In our measurements, we have used at least 400 independent configurations for each parameter set. The autocorrelation time is negligible except for the Polyakov loop. The Polyakov loop experiences, as expected, large autocorrelation times close to the phase transition. These are taken into account in the error estimates.
Specifically, the errors of $\langle|P|\rangle$ and of the Polyakov-loop susceptibility are obtained from a bootstrap over blocks of $\lceil 4\tau_{\rm int}\rceil$ consecutive measurements, where $\tau_{\rm int}$ is the integrated autocorrelation time determined on the same series, and the identical prescription is applied to every data set shown. In the transition region, several independent chains started from ordered and disordered configurations are combined for each parameter set.
%%%%%%%%%%%%%%%%%%%%%%%%%%%
%%%%%%%%%%%%%%%%%%%%%%%%%%%
\subsection*{Convergence of spatial and temporal plaquette expectation values}
%%%%%%%%%%%%%%%%%%%%%%%%%%%
%%%%%%%%%%%%%%%%%%%%%%%%%%%
In Fig.~\ref{fig:plaquette_convergence}, spatial and temporal plaquettes are plotted against $1/m^2=1/m_{\rm U(1)}^2$. For these plots, the lattice size is $8^3$ and the lattice spacing is $a_s=a=0.3$ for SU(2) and $0.2$ for SU(3). There is a smooth convergence to the value obtained from the Wilson action as the mass is sent to infinity.

Throughout, the extrapolations to the infinite-mass limit are performed as expansions in powers of $1/m^2$. This functional form is dictated by the heavy-scalar nature of the limit: the scalar fluctuations scale as $\langle(ag_d\phi)^2\rangle\sim 1/m^2$, so that $\langle\mathrm{Tr}(W-\textbf{1}_N)^2\rangle\sim 1/m^2$, and integrating out the heavy scalar generates corrections to the observables organized in powers of $1/m^2$.\footnote{This differs from the $m^{-1/2}$ of Sec.~3 because our Euclidean lattice has $a_tm\gg1$ for these data, $a_tm\approx3$--$19$: the time step is coarse compared with the oscillation period $1/m$, the kinetic term is negligible against the mass term, and $\langle\phi^{2}\rangle\simeq1/(a_tm^{2})$. The exact free lattice result, $\langle\phi^{2}\rangle=1/(m\sqrt{4+a_t^{2}m^{2}})$, interpolates between this and the Hamiltonian value $1/(2m)$ at $a_tm\ll1$.}
The variable $(am)^{-2}$ used for the $\gamma_{\rm c}$ analysis below is the same expansion parameter up to the fixed factor $a^2$, so the two analyses employ the same expansion; it is truncated at quadratic order for the $\gamma=0$ data, which lie very close to $1/m^2=0$, and at cubic order for the $\gamma_{\rm c}$ data, which extend to substantially larger $1/m^2$ (i.e.\ $am\sim\mathcal{O}(1)$).

Fig.~\ref{fig:plaquette_finite-T} shows the finite-temperature behavior. 
As we will see later using the Polyakov loop, the confinement-deconfinement transition takes place in the range of $a_t$ shown in this plot.
The gauge group is SU(3), and the lattice size is $4\times 32^2$. The horizontal axis is $a_t$, which is related to the temperature $T$ (in units of the coupling) by $T=\frac{1}{4a_t}$. The spatial lattice spacing is fixed to $a=0.2$.  We can see that $\langle\mathrm{Tr}(UUU^\dagger U^\dagger)\rangle_{\rm spatial}$ and $\langle\mathrm{Tr}(UUU^\dagger U^\dagger)\rangle_{\rm temporal}$ converge to the values in the Wilson action quickly at all values of $a_t$, suggesting a decoupling of scalars and the U(1) part from the SU($N$) gauge field.

\begin{figure}[htbp]
	\centering
    \scalebox{0.6}{\input{Fig_SU2_plaquette}}
    \scalebox{0.6}{\input{Fig_SU3_plaquette}}
    \hfill
	\caption{Spatial and temporal plaquette. [Left] SU(2), $8^3$ lattice, $a_t=a=0.3$.
    [Right] SU(3), $8^3$ lattice, $a_t=a=0.2$.
    Infinite-mass extrapolations by a quadratic function in $1/m^2$ from $m^2=250,\cdots,4000$ are shown at $1/m^2=0$. 
    The horizontal lines are the values obtained from the Wilson action. 
    For $\langle\mathrm{Tr}(UUU^\dagger U^\dagger)\rangle_{\rm spatial}$ and $\langle\mathrm{Tr}(UUU^\dagger U^\dagger)\rangle_{\rm temporal}$, the horizontal axis is slightly shifted so that the data points can be distinguished. 
    }
\label{fig:plaquette_convergence}
\end{figure}

\begin{figure}[htbp]
	\centering
    \scalebox{0.6}{\input{Fig_spatial_plaquette_vs_atgnuplot}}
    \scalebox{0.6}{\input{Fig_temporal_plaquette_vs_atgnuplot}}
    \hfill
\caption{$\langle\mathrm{Tr}(UUU^\dagger U^\dagger)\rangle_{\rm spatial}$ [Left] and $\langle\mathrm{Tr}(UUU^\dagger U^\dagger)\rangle_{\rm temporal}$ [Right] versus $a_t$, for SU(3) on a $4\times 32^2$ lattice with fixed spatial spacing $a_s=0.2$. Orbifold-lattice results at $m^2=m^2_{\rm U(1)}=100,\,300,\,1000$ are compared with the Wilson action. Each point is based on at least 400 configurations, separated by 150 HMC trajectories; errors are autocorrelation-corrected.
    }
\label{fig:plaquette_finite-T}
\end{figure}
%%%%%%%%%%%%%%%%%%%%%%%%%%%
%%%%%%%%%%%%%%%%%%%%%%%%%%%
\subsection*{Polyakov loop}
%%%%%%%%%%%%%%%%%%%%%%%%%%%
%%%%%%%%%%%%%%%%%%%%%%%%%%%
The Polyakov loop is constructed by taking a trace of the product of temporal links at each spatial point $(x,y)$ as
\begin{align}
    P_{x,y}
    =
    \frac{1}{N}\mathrm{Tr}\left(
    U_{t,\vec{n}=(1,x,y)}
    U_{t,\vec{n}=(2,x,y)}
    \cdots
    U_{t,\vec{n}=(n_t,x,y)}
    \right)\, . 
\end{align}
We take the spatial average for each configuration,
\begin{align}
    P = \frac{1}{n_xn_y}\sum_{x,y}P_{x,y}.
\end{align}
For simplicity, we call this quantity $P$ Polyakov loop in the following. 

A simple way to characterize the deconfinement transition in this theory is the breaking of $\mathbb{Z}_N$ center symmetry that acts on the Polyakov loop as $P\to e^{2\pi\mathrm{i}/N}P$. In the large-volume limit, a nonzero expectation value of $P$ indicates broken center symmetry. 
However, in a finite volume, tunnelings between different sectors $\mathbb{Z}_N$ take place and $\langle P\rangle$ vanishes even in the broken regime. Therefore, we use $\langle |P|\rangle$ instead. 

\begin{figure}[htbp]
	\centering
    \scalebox{0.6}{\includegraphics{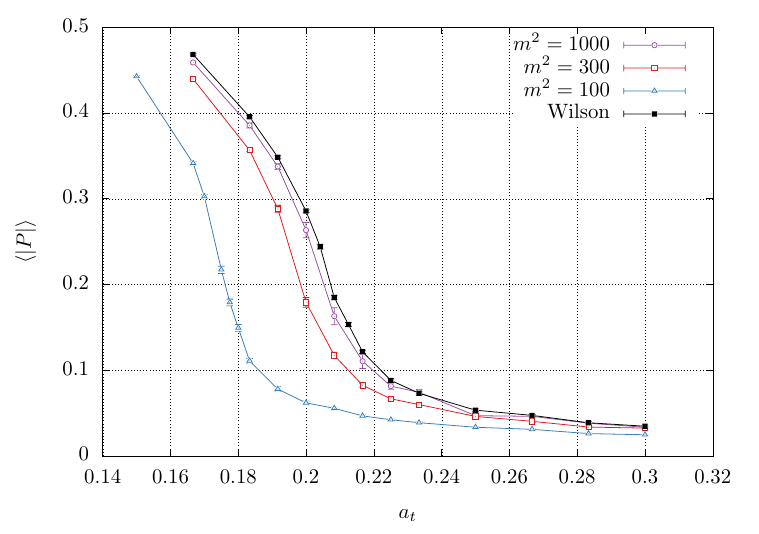}}
    \scalebox{0.6}{\includegraphics{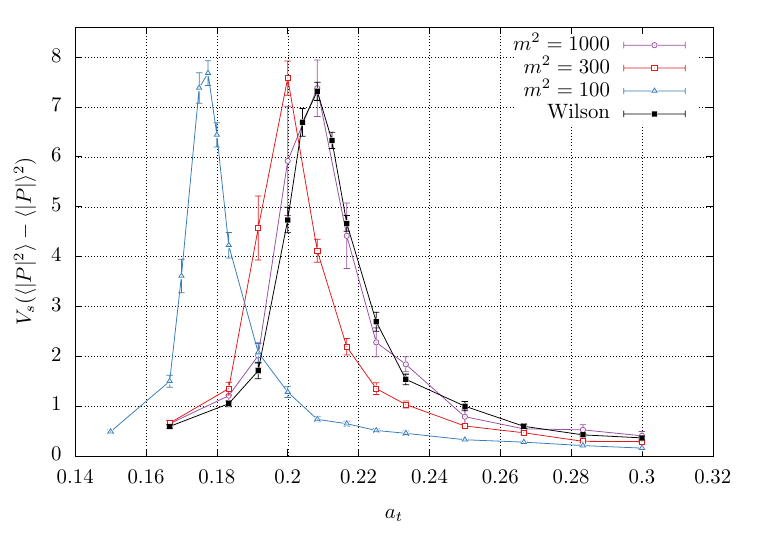}}
    \hfill
	\caption{[Left] $\langle|P|\rangle$ versus $a_t$. [Right] the Polyakov-loop susceptibility $\chi\equiv V_s\left(\langle |P|^2\rangle-\langle |P|\rangle^2\right)$, with spatial volume $V_s=n_xn_y$, versus $a_t$. SU(3), $4\times 32^2$ lattice, fixed $a_s=0.2$; orbifold results at $m^2=100,\,300,\,1000$ compared with the Wilson action. Based on the same configurations as Fig.~\ref{fig:plaquette_finite-T}, supplemented near the transition by additional independent chains started from ordered and disordered configurations. The rise of $\langle|P|\rangle$ and the peak of the susceptibility locate the deconfinement transition. Note that the transition of the orbifold theory moves towards the Wilson-action result as $m^2$ is increased.
	Rather large $m^2$-dependence is due to a shift of effective lattice spacing, as we will see from Fig.~\ref{fig:finite-T-Pol_gammac}. 
    }
\label{fig:finite-T-Pol}
\end{figure}

In Fig.~\ref{fig:finite-T-Pol}, $\langle |P|\rangle$ and the Polyakov-loop susceptibility $\chi \equiv V_s\left(\langle |P|^2\rangle - \langle |P|\rangle^2\right)$, with spatial volume, are shown. The increase in $\langle|P|\rangle$ and the peak of $\chi$ indicate the deconfinement transition. We observe convergence to the values of the Wilson action as the mass is sent to infinity, across a wide temperature region, including the phase transition. 
The same convergence holds for the distribution of $P$; see Fig.~\ref{fig:finite-T-Pol-dist}.
The residual deviation that remains at the smallest mass, $m^2=100$ --- the transition is located at a visibly smaller $a_t$ than for the Wilson action --- is revisited at the end of Sec.~\ref{sec:numerical_demonstration}, where the improvement term $\hat{H}_{\rm linear}$ is switched on; see Fig.~\ref{fig:finite-T-Pol_gammac}.

\begin{figure}
    \centering
    \includegraphics[width=.7\textwidth]{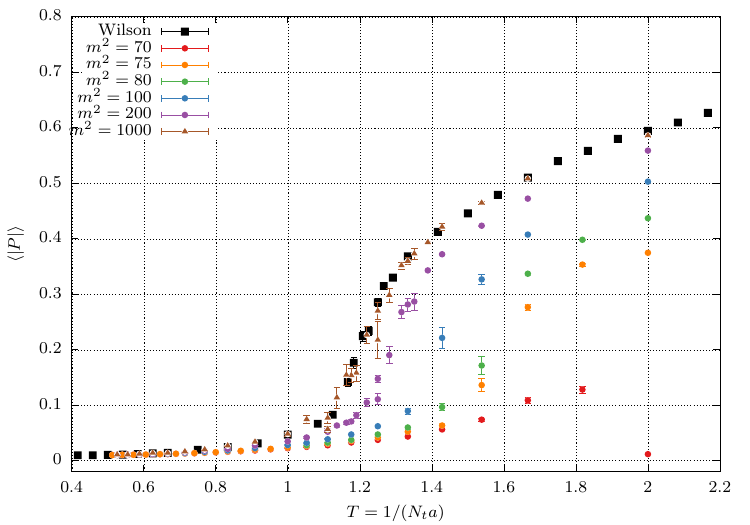}
	\caption{A parameter scan with the orbifold action of SU(3) Yang-Mills theory on a $4\times 32^2$ lattice. The lattice spacing is isotropic, $a=a_t$. For comparison, data from the Wilson action has been added; the Wilson action data were obtained with a heat bath algorithm, whereas all other data were computed with HMC.}
\label{fig:finit-T-Pol_iso}
\end{figure}

The simplest investigation of the phase transition in Yang-Mills theory usually considers isotropic lattices without a continuum extrapolation. In order to test this scenario, we have set the same lattice spacing in both directions $a=a_t$, which also determines the temperature. This is the same as a scan of the results as a function of the coupling constant. The results are shown in Fig.~\ref{fig:finit-T-Pol_iso} and illustrate the convergence to the Wilson action results. In this case we have also added some lower masses, which represent a stronger deviation from the Wilson data. The investigation of the lower mass regime is not the focus of the current paper, but will be investigated in a future publication.

\begin{figure}[htbp]
	\centering
    \scalebox{0.4}{\input{Fig_Wilson_Pol_distribution_SU3_at020}}
    \scalebox{0.4}{\input{Fig_Wilson_Pol_distribution_SU3_at025}}
    \scalebox{0.4}{\input{Fig_Wilson_Pol_distribution_SU3_at030}}
    \scalebox{0.4}{\input{Fig_Orbifold_Pol_distribution_SU3_M1000_at020}}
    \scalebox{0.4}{\input{Fig_Orbifold_Pol_distribution_SU3_M1000_at025}}
    \scalebox{0.4}{\input{Fig_Orbifold_Pol_distribution_SU3_M1000_at030}}
    \scalebox{0.4}{\input{Fig_Orbifold_Pol_distribution_SU3_M300_at020}}
    \scalebox{0.4}{\input{Fig_Orbifold_Pol_distribution_SU3_M300_at025}}
    \scalebox{0.4}{\input{Fig_Orbifold_Pol_distribution_SU3_M300_at030}}
    \scalebox{0.4}{\input{Fig_Orbifold_Pol_distribution_SU3_M100_at020}}
    \scalebox{0.4}{\input{Fig_Orbifold_Pol_distribution_SU3_M100_at025}}
    \scalebox{0.4}{\input{Fig_Orbifold_Pol_distribution_SU3_M100_at030}}    
    \hfill
	\caption{The distributions of $P$ from the Monte Carlo simulations. The horizontal and vertical axes are $\mathrm{Re}P$ and $\mathrm{Im}P$, respectively. SU(3), $4\times 32^2$ lattice, fixed $a_s=0.2$; orbifold results at $m^2=100,\,300,\,1000$ compared with the Wilson action. Based on the same configurations as Fig.~\ref{fig:finite-T-Pol}.
    }
\label{fig:finite-T-Pol-dist}
\end{figure}

%%%%%%%%%%%%%%%%%%%%%%%%%%%
%%%%%%%%%%%%%%%%%%%%%%%%%%%
\subsection*{Convergence of $\langle\mathrm{Tr}(W-\textbf{1}_N)^2\rangle$ and $\langle\det U\rangle$}
%%%%%%%%%%%%%%%%%%%%%%%%%%%
%%%%%%%%%%%%%%%%%%%%%%%%%%%
In Fig.~\ref{fig:TrW_Re_det_U}, $\langle\mathrm{Tr}(W-\textbf{1}_N)^2\rangle$ and $\langle\mathrm{Re}(\det U)\rangle$ are plotted for SU(2) and SU(3), $8^3$ lattice, taking horizontal axis $1/m^2=1/m^2_{\rm U(1)}$. We can see the convergence to 0 and 1 as the mass is sent to infinity.  

In Fig.~\ref{fig:TrW_Re_det_U_finite-T}, the behaviors of these quantities at finite temperature are shown for SU(3). The approach to the infinite-mass limit can be seen at any temperature in the plots. 

\begin{figure}[htbp]
	\centering
\scalebox{0.6}{\input{Fig_TrW_N2t8s8}}
\scalebox{0.6}{\input{Fig_Re_detU_N2t8s8}}
\scalebox{0.6}{\input{Fig_TrW_N3t8s8}}
\scalebox{0.6}{\input{Fig_Re_detU_N3t8s8}}   
    \hfill
	\caption{$\mathrm{Tr}(W-\textbf{1}_N)^2$ (left column) and $\mathrm{Re}(\det U)$ (right column). 
    [Top] SU(2), $8^3$ lattice, $a_t=a=0.3$. 
    [Bottom] SU(3), $8^3$ lattice, $a_t=a=0.2$.
    Infinite-mass extrapolations by a quadratic function of $1/m^2$ from $m^2=250,\cdots,4000$ are shown at $1/m^2=0$. 
    }
\label{fig:TrW_Re_det_U}
\end{figure}

\begin{figure}[htbp]
	\centering
\scalebox{0.6}{\input{Fig_TrW_finite-T}}
\scalebox{0.6}{\input{Fig_Re_detU_finite-T}}   
    \hfill
\caption{[Left] $\mathrm{Tr}(W-\textbf{1}_N)^2$ versus $a_t$. [Right] $\mathrm{Re}(\det U)$ versus $a_t$.
    SU(3), $4\times 32^2$ lattice, fixed $a_s=0.2$; orbifold results at $m^2=100,\,300,\,1000$. Based on the same configurations as Fig.~\ref{fig:plaquette_finite-T}.
    }
\label{fig:TrW_Re_det_U_finite-T}
\end{figure}
%%%%%%%%%%%%%%%%%%%%%%%%%%%
%%%%%%%%%%%%%%%%%%%%%%%%%%%
\subsection{Improved extrapolation to the Kogut-Susskind limit}
%%%%%%%%%%%%%%%%%%%%%%%%%%%
%%%%%%%%%%%%%%%%%%%%%%%%%%%
So far, we demonstrated the convergence to the Kogut-Susskind limit with fixed \textit{bare} lattice spacing. The \textit{effective} lattice spacing was different from the bare lattice spacing when the mass parameters are finite. 

Now, we study the Kogut-Susskind limit at fixed effective lattice spacing. Specifically, we introduce $\hat{H}_{\rm linear}$. (As another option, we could also tune bare lattice spacing to keep the effective lattice spacing fixed~\cite{Bergner:2026duh}.) At each value of $m^2$, we find $\gamma=\gamma_{\rm c}$ where $\langle\mathrm{Tr}(W-1)\rangle$ becomes zero, and then, study the $m^2$-dependence of several quantities at $\gamma=\gamma_{\rm c}$. 
%%%%%%%%%%%%%%%%%%%%%%%%
%%%%%%%%%%%%%%%%%%%%%%%%
%\newpage
\subsection*{Determination of $\gamma_{\rm c}$}
\begin{figure}[H]
    \centering
    \begin{subfigure}{0.5\textwidth}
        \centering
    \includegraphics[width=1.0\linewidth]{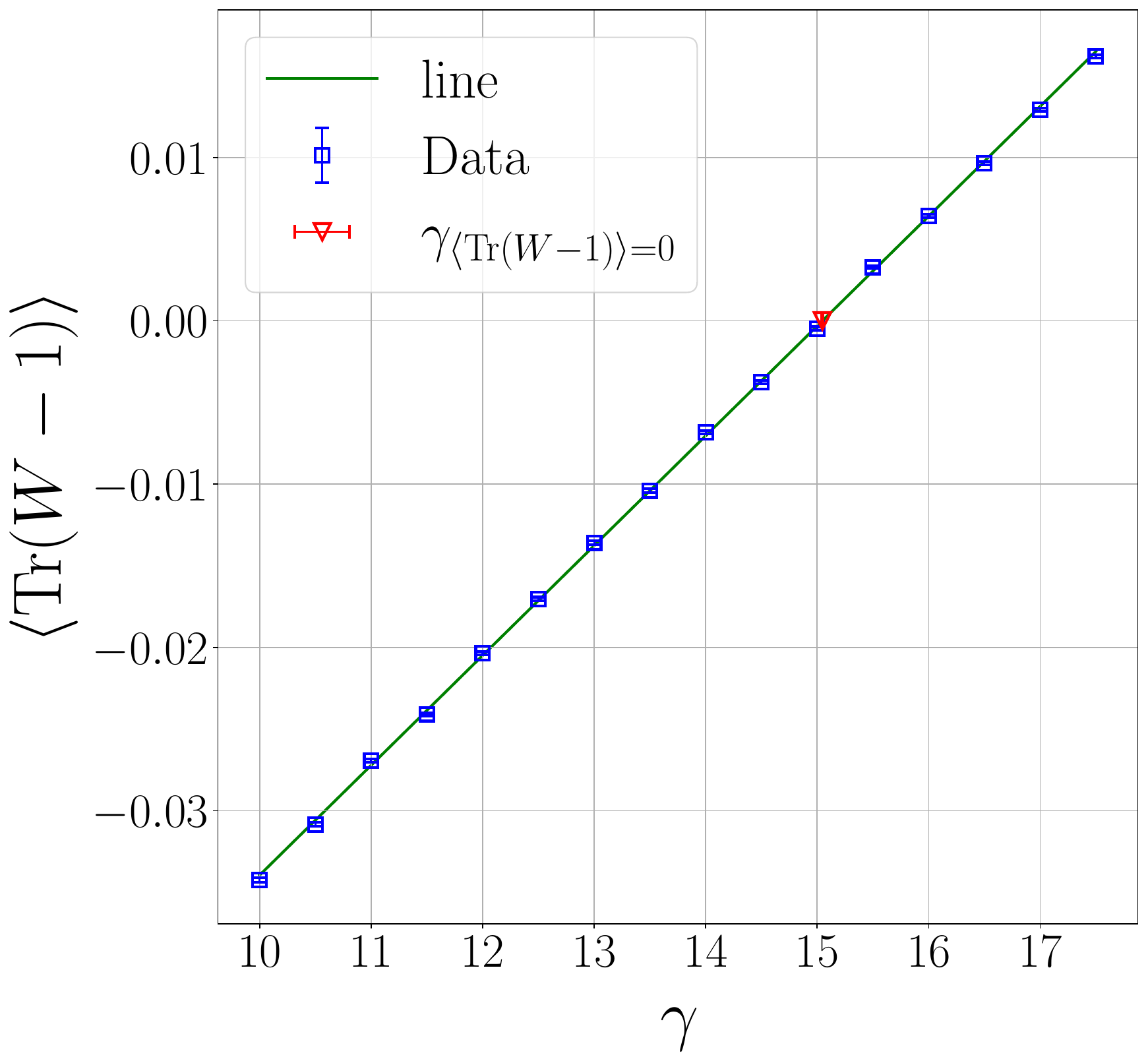}
    \end{subfigure}
    \caption{ 
    Fit of $\langle \mathrm{Tr}(W - \mathbf{1}_N) \rangle$  values versus $\gamma $, $8^3$ lattice, 
    SU(3), $a=a_t=0.2$, $m^2=125$.
} \label{fig:example_fit_TrW_1}
\end{figure}

%%%%%%%%%%%%%%%%%%%%%%%%
%%%%%%%%%%%%%%%%%%%%%%%%
To determine $ \gamma_{\rm c}$ at each $m^2$, we perform simulations at several values of $\gamma$, fit $\langle \mathrm{Tr}(W - \mathbf{1}_N) \rangle$ as a function of $\gamma$, simply using $\langle \mathrm{Tr}(W - \mathbf{1}_N) \rangle = A\gamma+B$. An example is shown in Fig.~\ref{fig:example_fit_TrW_1}. Tables~\ref{tab:gamma_crit_SU2_am} and~\ref{tab:gamma_crit_SU3_am} summarize the results for SU(2) and SU(3) theories, respectively, on an $8^3$ lattice with $a=a_t=0.3$ for SU(2) and $a=a_t=0.2$ for SU(3).

\begin{table}[htbp]
\centering
\begin{subtable}[t]{0.48\textwidth}
\centering
\footnotesize
%Table for SU2  a_t=a_s = 0.3
\begin{tabular}{|c||c|c|}
\hline
$m^2$ & $\gamma_{\rm c}$  & $a \times m$ \\
\hline\hline
5   & 9.746680 & 0.670820 \\
9   & 7.730493 & 0.900000 \\
11  & 7.671210 & 0.994987 \\
13  & 7.623557 & 1.081665 \\
16  & 7.575180 & 1.200000 \\
22  & 7.502837 & 1.407125 \\
25  & 7.476063 & 1.500000 \\
28  & 7.446263 & 1.587451 \\
32  & 7.408923 & 1.697056 \\
36  & 7.386837 & 1.800000 \\
40  & 7.354473 & 1.897367 \\
44  & 7.346650 & 1.989975 \\
50  & 7.315410 & 2.121320 \\
75  & 7.243050 & 2.598076 \\
100 & 7.194417 & 3.000000 \\
125 & 7.161960 & 3.354102 \\
150 & 7.140743 & 3.674235 \\
175 & 7.106953 & 3.968627 \\
200 & 7.094080 & 4.242641 \\
225 & 7.087750 & 4.500000 \\
250 & 7.073507 & 4.743416 \\
275 & 7.065953 & 4.974937 \\
300 & 7.046990 & 5.196152 \\
\hline
\end{tabular}
\caption{SU(2), $a=a_t=0.3$.}
\label{tab:gamma_crit_SU2_am}
\end{subtable}
\hfill
\begin{subtable}[t]{0.48\textwidth}
\centering
\footnotesize
\begin{tabular}{|c||c|c|}
\hline
$m^2$ & $\gamma_{\rm c}$ & $a \times m$ \\
\hline\hline
5.0   & 16.315680 & 0.447214 \\
9.0   & 18.561845 & 0.6      \\
16.0  & 17.933675 & 0.8      \\
20.0  & 17.726980 & 0.894427 \\
25.0  & 17.467360 & 1.0      \\
30.0  & 17.283905 & 1.095445 \\
36.0  & 17.060590 & 1.2      \\
42.0  & 16.854545 & 1.296148 \\
49.0  & 16.629630 & 1.4      \\
50.0  & 16.586855 & 1.414214 \\
56.0  & 16.438995 & 1.496663 \\
64.0  & 16.222980 & 1.6      \\
72.0  & 16.027750 & 1.697056 \\
75.0  & 15.951550 & 1.732051 \\
81.0  & 15.837810 & 1.8      \\
90.0  & 15.643905 & 1.897367 \\
100.0 & 15.459455 & 2.0      \\
125.0 & 15.047240 & 2.236068 \\
150.0 & 14.676130 & 2.449490 \\
175.0 & 14.389375 & 2.645751 \\
200.0 & 14.131770 & 2.828427 \\
225.0 & 13.930280 & 3.000    \\
250.0 & 13.718430 & 3.162    \\
275.0 & 13.553465 & 3.317    \\
300.0 & 13.397350 & 3.464    \\
\hline
\end{tabular}
\caption{SU(3), $a=a_t=0.2$.}
\label{tab:gamma_crit_SU3_am}
\end{subtable}
\caption{The values of $\gamma_{\rm c}$ used to demonstrate the improved extrapolation to the infinite-mass limit with fixed effective lattice spacing, and the corresponding $m^2$ and $a\times m$. The lattice size is $8^3$. [left] SU(2) at $a=a_t=0.3$, [right] SU(3) at $a=a_t=0.2$.}
\label{tab:gamma_crit_am}
\end{table}

%%%%%%%%%%%%%%%%%%%%%%%%
%%%%%%%%%%%%%%%%%%%%%%%%
\subsection*{$m^2$-dependence at $\gamma=\gamma_{\rm c}$ }
\begin{figure}[H]
    \centering
    \begin{subfigure}{0.49\textwidth}
        \centering
    \includegraphics[width=1.01\linewidth]{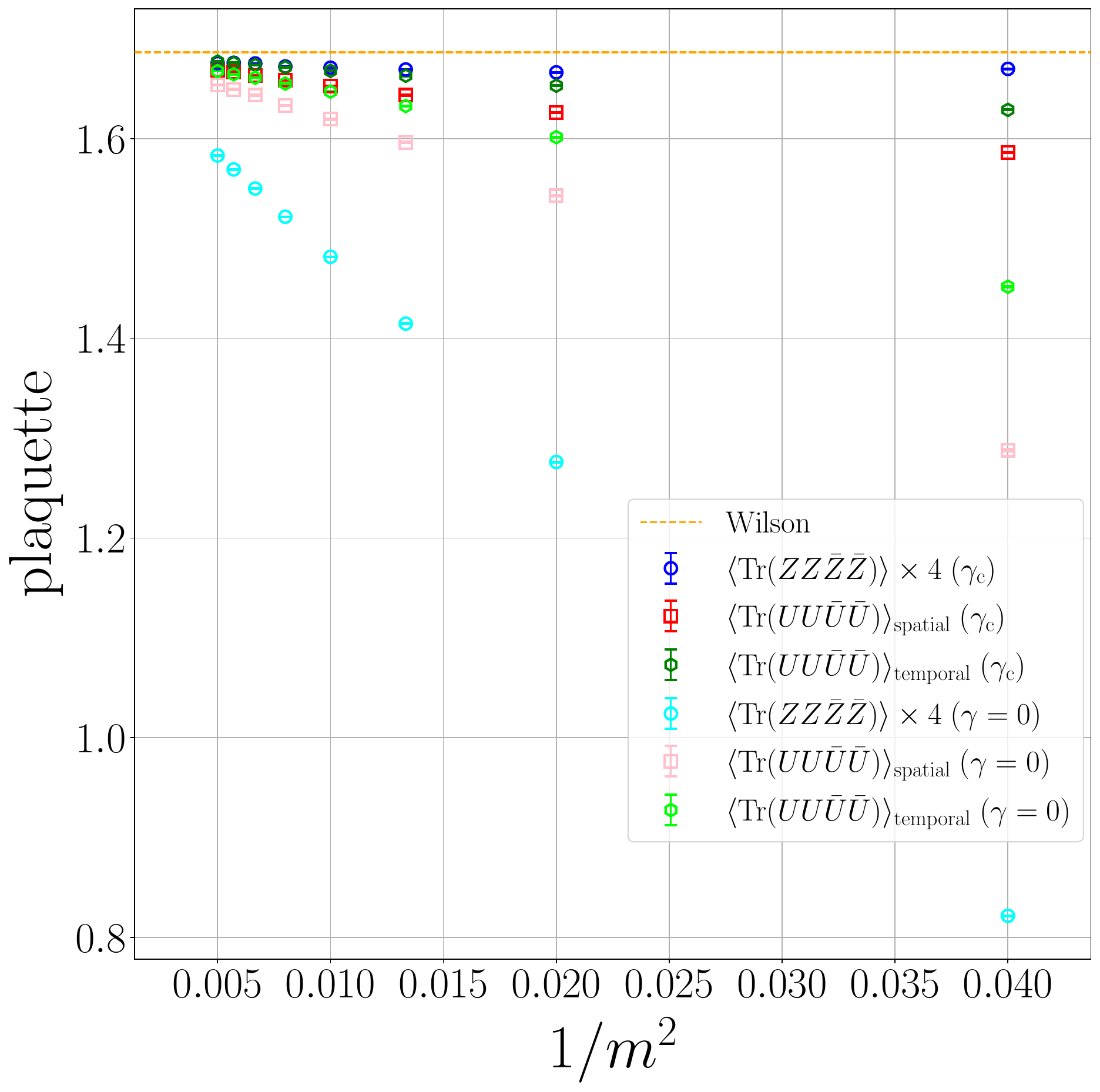}
     %   \label{fig:sub2__}
\end{subfigure}
    \hfill
    \begin{subfigure}{0.49\textwidth}
        \centering
    \includegraphics[width=1.0\linewidth]{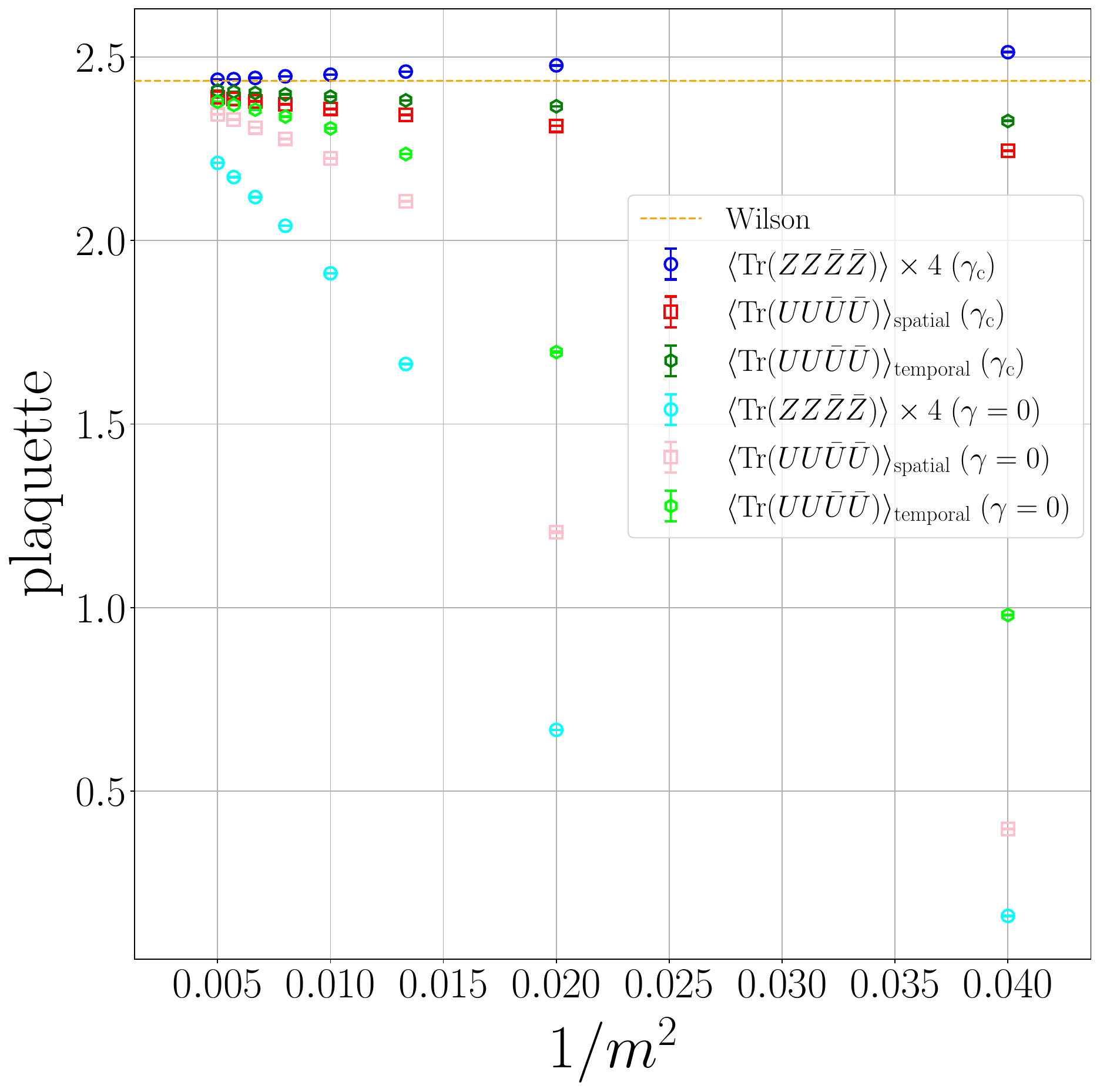}
    \end{subfigure}
    \caption{
    Plaquette values versus $1/m^2$ at $\gamma_{\rm c}$, $8^3$ lattice, 
    [\textbf{Left}] SU(2), $a=a_t=0.3$, and 
    [\textbf{Right}] SU(3), $a=a_t=0.2$.
} \label{fig:plaquettes_at_gamma_c}
\end{figure}

%%%%%%%%%%%%%%%%%%%%%%%%
%%%%%%%%%%%%%%%%%%%%%%%%
We performed Monte Carlo simulations at the values of $\gamma_{\rm c}$ summarized in Tables~\ref{tab:gamma_crit_SU2_am} and~\ref{tab:gamma_crit_SU3_am}.
Plaquette values are shown in Fig.~\ref{fig:plaquettes_at_gamma_c}, taking $1/m^2$ the horizontal axis. We can see a significant improvement from $\gamma=0$. This shows that the most of the ``deviation from the Kogut-Susskind limit" observed at $\gamma=0$ merely came from a shift of lattice spacing from the input value. 

\begin{figure}[H]
    \centering
    \begin{subfigure}{0.49\textwidth}
        \centering
    \includegraphics[width=1.01\linewidth]{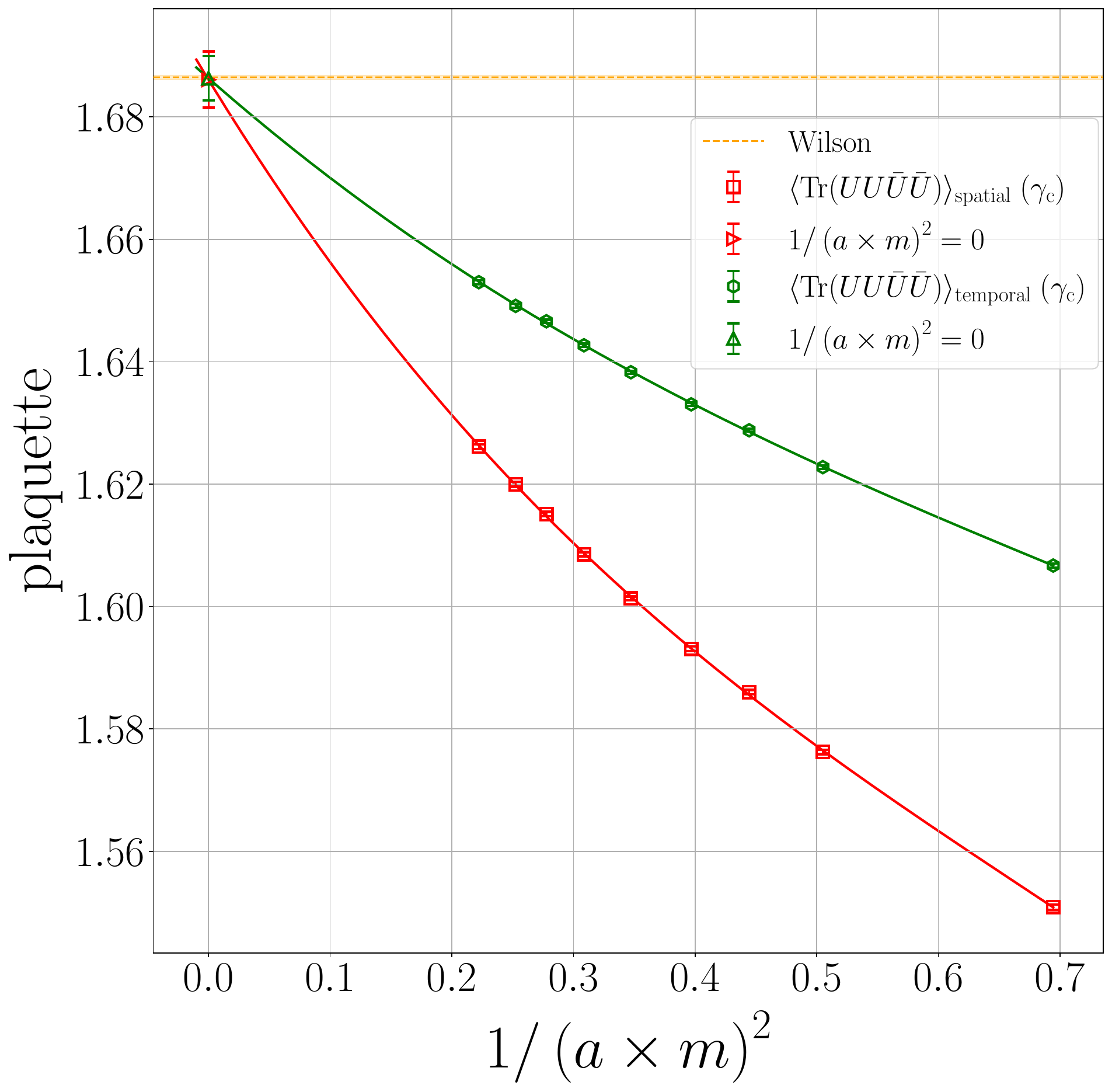}
     %   \label{fig:sub2__}
\end{subfigure}
    \hfill
    \begin{subfigure}{0.49\textwidth}
        \centering
    \includegraphics[width=1.01\linewidth]{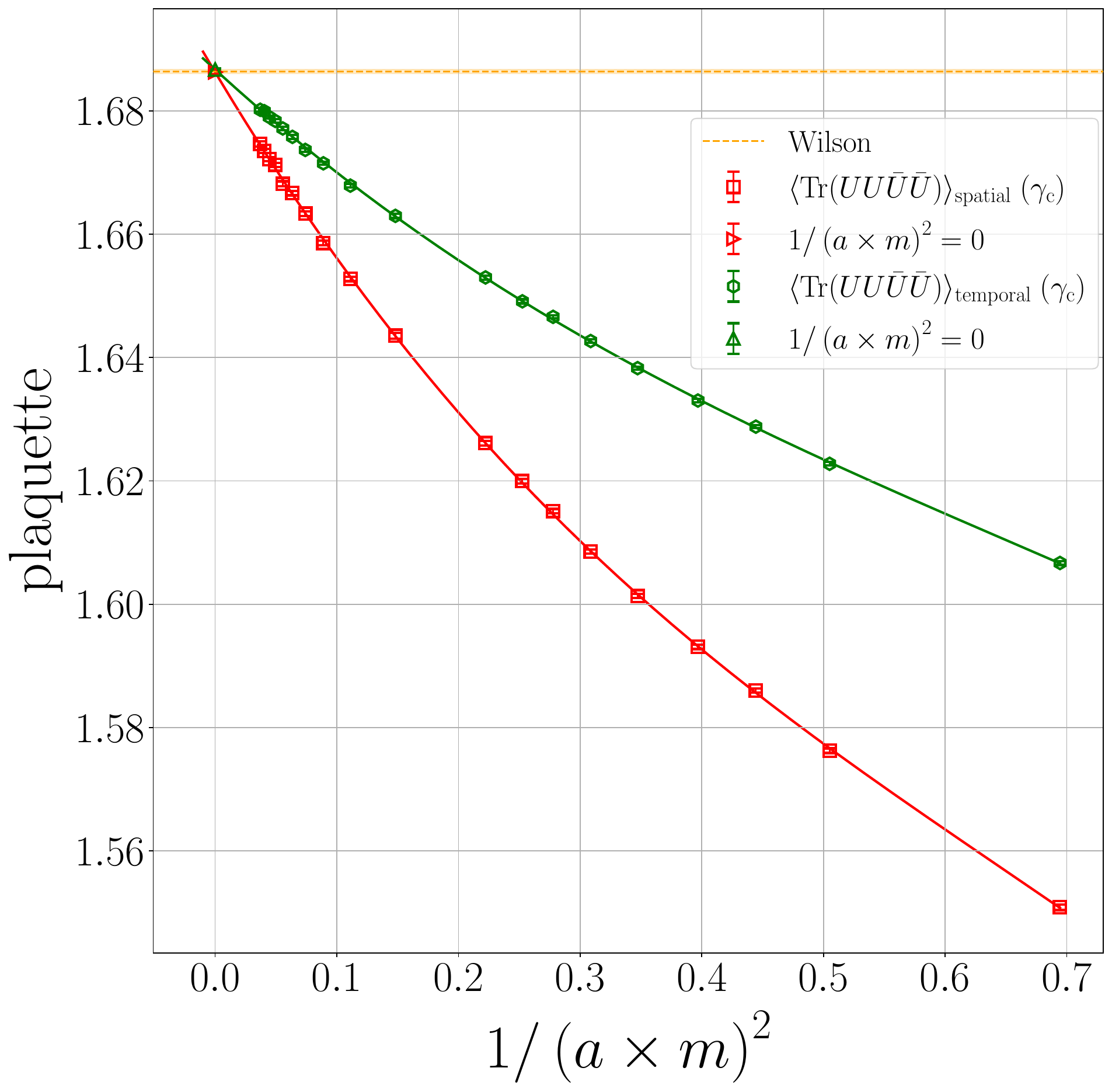}
    \end{subfigure}
    \caption{ 
Extrapolation of SU(2) plaquette values toward the KS limit using simulations at $\gamma_{\rm c}$ with $a , m \sim \mathcal{O}(1)$ on an $8^3$ lattice with $a = a_t = 0.3$. [\textbf{Left}] Data in the range $a \times m = [1.2, 2.1]$, and [\textbf{Right}] Data in the range $a \times m = [1.2, 5.2]$. The fit ansatz is a cubic polynomial of $(am)^{-2}$ (equivalently, of $1/m^2$; see text).
} \label{fig:Pla_am_SU2}
\end{figure}

\begin{figure}[H]
    \centering
    \begin{subfigure}{0.49\textwidth}
        \centering
    \includegraphics[width=1.05\linewidth]{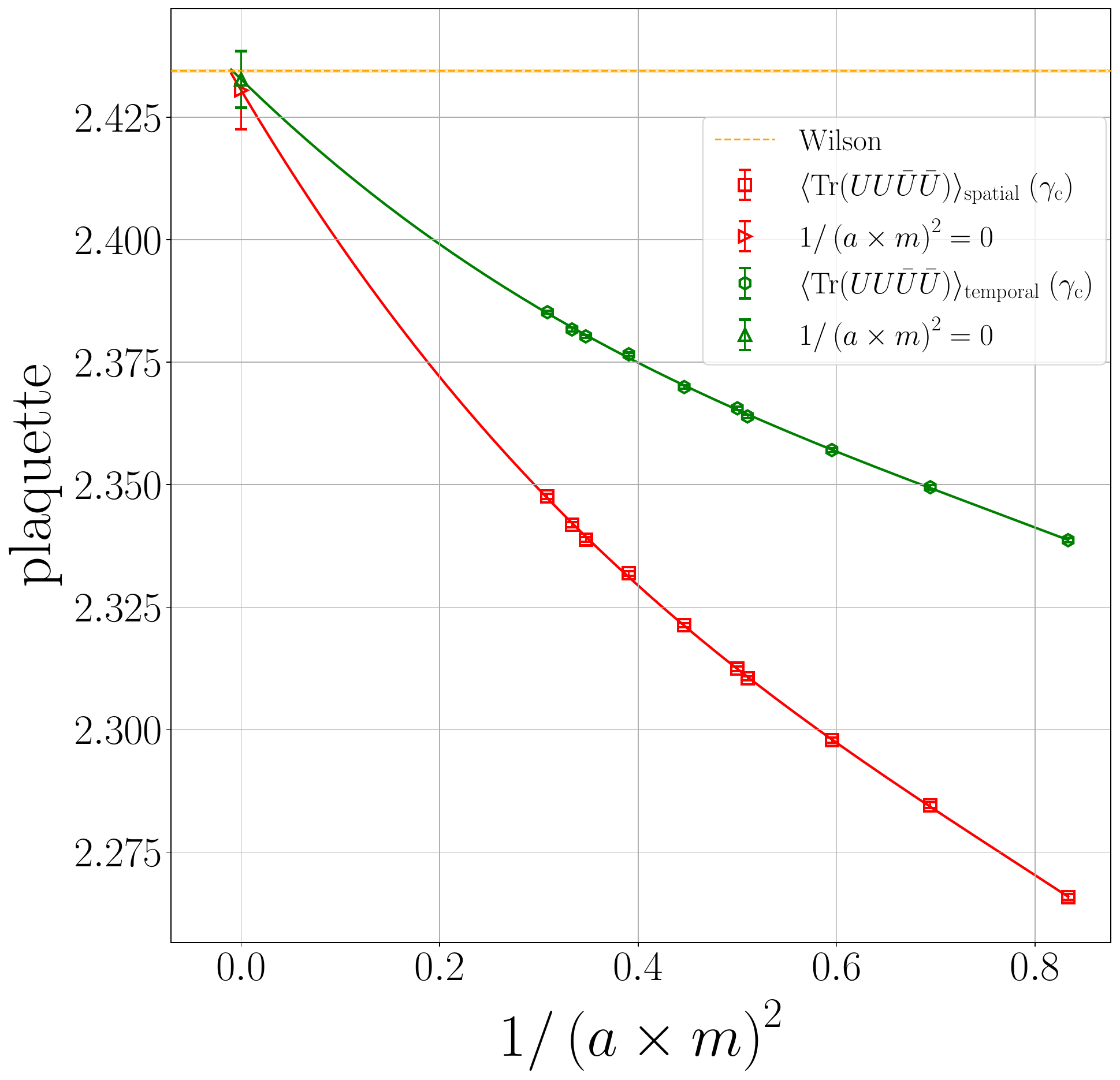}
     %   \label{fig:sub2__}
\end{subfigure}
    \hfill
    \begin{subfigure}{0.49\textwidth}
        \centering
    \includegraphics[width=1.03\linewidth]{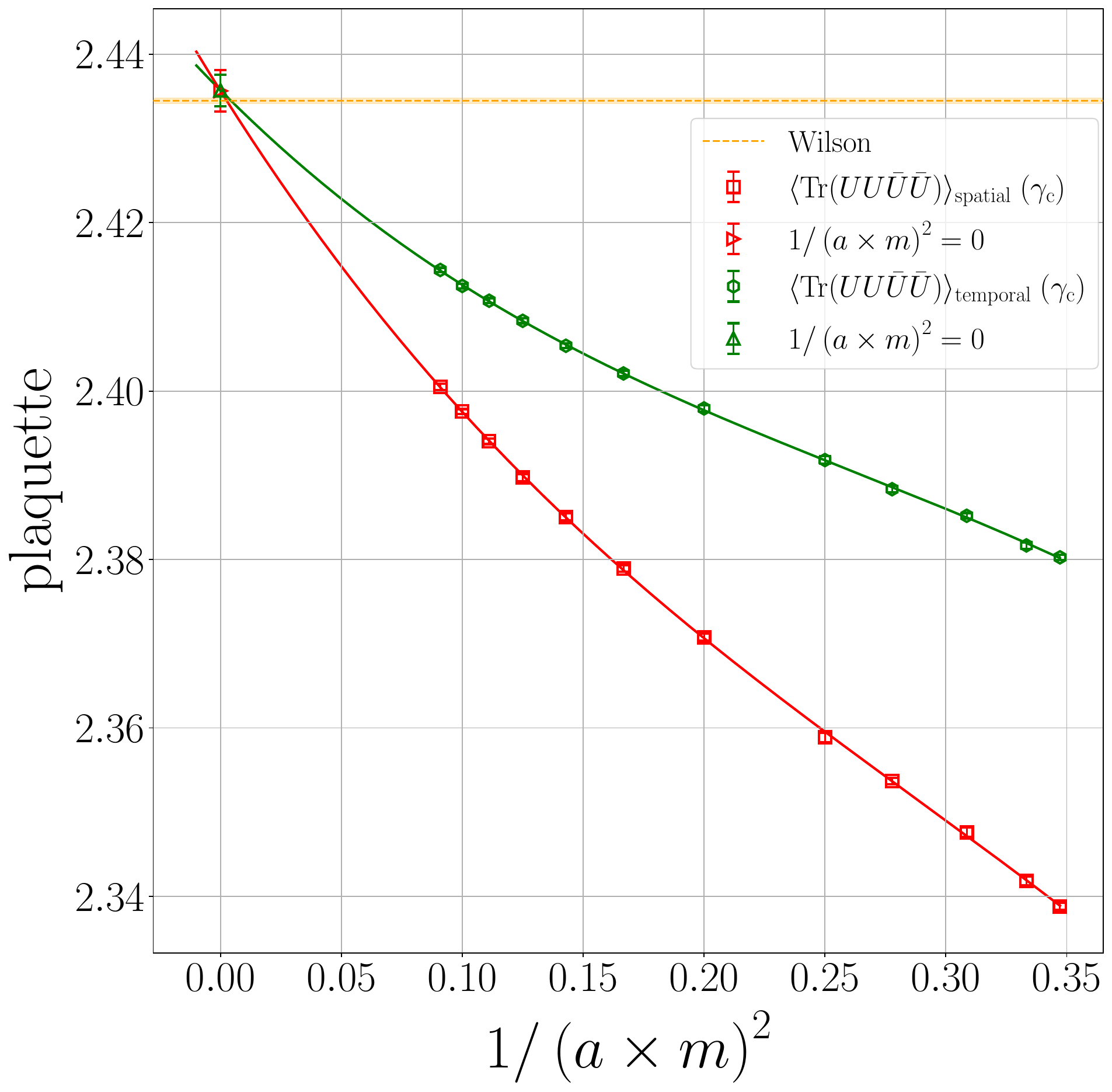}
    \end{subfigure}
    \caption{ 
Extrapolation of SU(3) plaquette values toward the KS limit using simulations at $\gamma_{\rm c}$ with $a , m \sim \mathcal{O}(1)$ on an $8^3$ lattice with $a = a_t = 0.2$. [\textbf{Left}] Data in the range $a \times m = [1.1, 1.8]$, and [\textbf{Right}] Data in the range $a \times m = [1.7, 3.3]$. The fit ansatz is a cubic polynomial of $(am)^{-2}$ (equivalently, of $1/m^2$; see text).
} \label{fig:Pla_am_SU3}
\end{figure}

Furthermore, the effectiveness of tuning $\gamma_{\rm c}$ can be better appreciated by examining how the plaquette values obtained at $\gamma_{\rm c}$ for different values of $a \times m \sim \mathcal{O}(1)$ --- shown in Table~\ref{tab:gamma_crit_SU2_am} for SU(2) and in Table~\ref{tab:gamma_crit_SU3_am} for SU(3) --- can be used to extrapolate their values to the KS limit ($m\to\infty$). 
These extrapolations, based exclusively on plaquette values obtained from simulations at $\gamma_{\rm c}$ with $a \times m \sim \mathcal{O}(1)$, are shown in Fig.~\ref{fig:Pla_am_SU2} for SU(2) at $a_s = a_t = 0.3$, and in Fig.~\ref{fig:Pla_am_SU3} for SU(3) at $a_s = a_t = 0.2$. Therefore, as expected, the KS limit can be evaluated using $m\sim a^{-1}$, without introducing a new scale other than $a^{-1}$. We need only the same energy scales as the KS formulation, i.e., the cutoff scale $a^{-1}$, and actual physical scales of interest such as the glueball mass. 

Since the action depends on the scalar mass only through $m^{2}$, observables at finite lattice spacing are even functions of $m$, and we extrapolate in powers of $1/m^{2}$, equivalently of $(am)^{-2}$. The series need not converge over the whole range of $a\times m$ used here, so the fits of Figs.~\ref{fig:Pla_am_SU2} and \ref{fig:Pla_am_SU3} should be read as finite-range fits; that the two nearly disjoint ranges in each figure lead to the same extrapolated value, in agreement with the Wilson action, supports this description. Odd powers of $1/m$ would appear only in the Hamiltonian limit $a_t\to0$ taken at fixed $m$, where $\langle\phi^{2}\rangle\to1/(2m)$; all data shown here are at $a_t=a$. In any case, the infinite-mass limit is optional, and no physical conclusion depends on this extrapolation.

%%%%%%%%%%%%%%%%%%%%%%%%
%%%%%%%%%%%%%%%%%%%%%%%%
%%% NEW SUBSECTION (v4) %%%
\subsection*{The deconfinement transition at $\gamma=\gamma_{\rm c}$}\label{sec:gammac_finiteT}
%%%%%%%%%%%%%%%%%%%%%%%%
%%%%%%%%%%%%%%%%%%%%%%%%
So far, the improvement term has been tested with ultraviolet observables at zero temperature. It is natural to ask whether fixing the effective lattice spacing also improves an infrared quantity. The confinement-deconfinement transition provides such a test, and a demanding one: as noted in Sec.~\ref{sec:numerical_demonstration}, the location of the transition is the largest deviation from the Kogut-Susskind limit that we observe at $m^2=100$.

We have therefore repeated the finite-temperature scan of Fig.~\ref{fig:finite-T-Pol} at $m^2=m^2_{\rm U(1)}=100$ with the improvement term switched on. The coefficient is \textit{not} re-tuned at each temperature. What is held fixed across the scan is the combination $a_t\gamma$ that multiplies $\sum_{\vec{n}}\sum_{j}\mathrm{Tr}\left(Z_{j,\vec{n}}\bar{Z}_{j,\vec{n}}\right)$ in \eqref{eq:deltaS_orbifold}, at the value $a_t\gamma=3.09$ corresponding to $\gamma=\gamma_{\rm c}=15.46$ of Table~\ref{tab:gamma_crit_SU3_am} at the isotropic point $a_t=a=0.2$.
The product is the natural quantity to hold fixed from a strong coupling perspective. For each link, the departure of $W_{j,\vec{n}}$ from $\mathbf{1}_N$ that this term compensates originates from the flat integration measure for $Z_{j,\vec{n}}$, which carries no factor of $a_t$ (a detailed discussion will be given elsewhere).
In this way one and the same term is added to the action at every temperature, and the comparison introduces no additional temperature-dependent parameter.

This prescription is justified a posteriori by the tuning condition itself, which remains well satisfied over the entire scan. At $a_t=0.2$, where $\gamma_{\rm c}$ was determined, we find $\langle\mathrm{Tr}(W-\textbf{1}_N)\rangle=0.0001(1)$, i.e.\ the condition is met on the $4\times 32^2$ lattice with the value determined independently on the $8^3$ lattice. Away from that point the residual grows only slowly and monotonically, staying below $0.011$ in absolute value over the whole range $0.167\leq a_t\leq 0.30$, to be compared with $0.10$--$0.17$ at $\gamma=0$; the linear term removes between $90\%$ and $99.9\%$ of the departure of $W_{j,\vec{n}}$ from $\textbf{1}_N$ across the scan. The quantity $\langle\mathrm{Tr}(W-\textbf{1}_N)^2\rangle$ of Fig.~\ref{fig:TrW_Re_det_U_finite-T} responds far less: it decreases by less than $3\%$ over the same range (from $0.188$ to $0.184$ at $a_t=0.2$).

\begin{figure}[htbp]
	\centering
    \scalebox{0.6}{\includegraphics{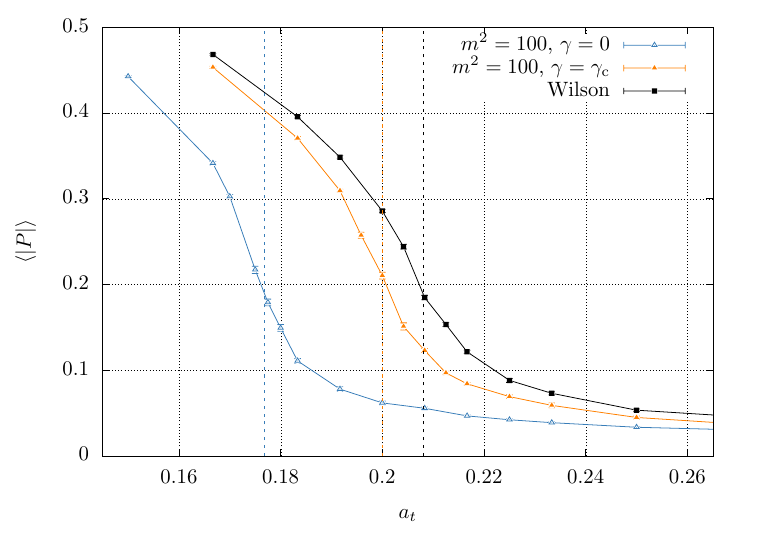}}
    \scalebox{0.6}{\includegraphics{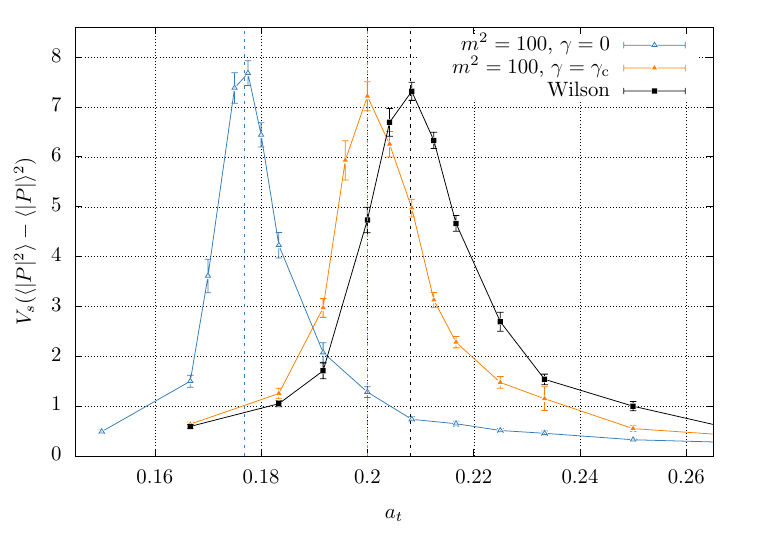}}
    \hfill
	\caption{Effect of the improvement term on the deconfinement transition. [Left] $\langle|P|\rangle$ versus $a_t$; [Right] the Polyakov-loop susceptibility $\chi=V_s\left(\langle|P|^2\rangle-\langle|P|\rangle^2\right)$ versus $a_t$. SU(3), $4\times 32^2$ lattice, fixed $a_s=0.2$, $m^2=m^2_{\rm U(1)}=100$ at $\gamma=0$ and at $\gamma=\gamma_{\rm c}$, compared with the Wilson action. The $\gamma=0$ and Wilson data are those of Fig.~\ref{fig:finite-T-Pol}. The vertical lines mark the susceptibility peak positions obtained from weighted parabola fits, $a_t=0.17686(18)$, $0.19997(99)$ and $0.20815(23)$ for $\gamma=0$, $\gamma=\gamma_{\rm c}$ and the Wilson action, respectively. Each chain consists of about $400$ measurements separated by $150$ HMC trajectories; at the values of $a_t$ closest to the peak, four to five independent chains with ordered and disordered starts are combined. Errors are obtained as described in Sec.~\ref{sec:numerical_demonstration}.
    }
\label{fig:finite-T-Pol_gammac}
\end{figure}

The effect on the transition is shown in Fig.~\ref{fig:finite-T-Pol_gammac}. At every $a_t$ the improved $\langle|P|\rangle$ lies between the $\gamma=0$ and the Wilson curve, with no crossings, and the susceptibility peak moves from
\begin{align}
a_t^{\gamma=0}=0.17686(18)
\qquad\text{to}\qquad
a_t^{\gamma=\gamma_{\rm c}}=0.19997(99)\, ,
\end{align}
to be compared with $a_t^{\rm Wilson}=0.20815(23)$; the peak positions are obtained from weighted parabola fits to the five points closest to the maximum. This means that the improvement removes about $74\%$ of the shift of the transition. Equivalently, in units of the coupling, the transition temperature $T=1/(4a_t)$ moves from $1.4134(14)$ to $1.2502(62)$, against $1.2011(13)$ for the Wilson action.

%%%%%%%%%%%%%%%%%%%%%%%%%%%
%%%%%%%%%%%%%%%%%%%%%%%%%%%
\section{Conclusions and outlook}\label{sec:conclusion}
%%%%%%%%%%%%%%%%%%%%%%%%%%%
%%%%%%%%%%%%%%%%%%%%%%%%%%%
A simple and universal framework~\cite{Halimeh:2024bth} applicable to a wide class of theories, including the orbifold lattice, leads to a significant speedup compared to more complex formulations such as the Kogut-Susskind Hamiltonian~\cite{Hanada:2025yzx}. Our work demonstrates that the Kogut-Susskind Hamiltonian can be obtained as a special limit of the orbifold lattice, allowing us to leverage this property to achieve exponential speedup in quantum simulations. Specifically, we can achieve exponential speedup in quantum simulation of the Kogut-Susskind Hamiltonian by performing quantum simulations of the orbifold lattice Hamiltonian at several values of mass parameters and extrapolating the results to the infinite-mass limit. Although the final result reproduces that of the Kogut-Susskind Hamiltonian, the simulations utilize the orbifold lattice Hamiltonian, thereby enabling us to implement the universal framework that delivers exponential speedup. Note that there is no need to introduce a new energy scale; as a simple dimensional analysis suggests, we can evaluate the Kogut-Susskind limit from bare scalar mass of order of the inverse lattice spacing, once the improvement term $\hat{H}_{\rm linear}$ tuned to $\gamma=\gamma_{\rm c}$ is included. Without this term, we used $am\gg1$ for the unimproved extrapolation.
The same term also improves an infrared quantity: at $m^2=100$ it removes a large part  of the shift of the deconfinement transition of the $(2+1)$-d SU(3) theory relative to the Wilson action, leaving a small residual that we have not attempted to remove.\footnote{If a nontrivial dynamics sets in only for the orbifold lattice, and not to Kogut-Susskind, in such a way that the quantum simulations become infeasible, then such an exponential speedup could be lost. However, we do not see an immediate reason for such a scenario, given that the former is simply a system of Yang-Mills theory coupled to scalars. If there were any issue for the orbifold lattice, the Kogut-Susskind formulation would also fail. 
}$^,$\footnote{There may be a better extrapolation ansatz even at $\gamma=0$ that takes into account the difference between the bare and effective lattice spacings.}

We emphasize that the mechanism demonstrated here is not restricted to pure Yang-Mills theory. The orbifold lattice formulation of QCD -- including quarks in the fundamental representation, in both the action and the Hamiltonian -- was constructed in ref.~\cite{Bergner:2024qjl}. There the noncompact link variable $Z_{j,\vec{n}}=W_{j,\vec{n}}U_{j,\vec{n}}$ reduces to the ordinary unitary link $U_{j,\vec{n}}$ as the scalar mass parameters $m^2$ and $m^2_{\rm U(1)}$ are sent to infinity (so that $W_{j,\vec{n}}\to\textbf{1}_N$ and $\det U_{j,\vec{n}}\to 1$) through the same type of term $\Delta\hat{H}$ (or $\Delta S_{\rm orbifold}$), and standard Wilson-fermion lattice QCD is recovered. Since this decoupling of the scalars is independent of the matter sector, the infinite-mass limit studied in the present work -- and hence the route to the Kogut-Susskind Hamiltonian and the associated exponential speedup -- carries over directly to full QCD. In particular, nothing in the limiting mechanism relies on the absence of matter, so the demonstration presented here should be read as evidence for the viability of the approach in QCD, not merely in pure gauge theory.

Note that several terms in the orbifold Hamiltonian \eqref{eq:Hamiltonian_orbifold} can be dropped without altering the infinite-mass limit. If we are interested only in this limit, it would be better to remove these terms and make the quantum circuits shorter. More generally, we might be able to use such flexibility in the choice of the Hamiltonian to connect the Kogut-Susskind limit and a simple model smoothly, allowing simpler extrapolations to the Kogut-Susskind limit. Furthermore, such an approach would enable adiabatic state preparation protocols with reduced computational overhead. We plan to explore this strategy using conventional lattice simulation methods on classical computers in forthcoming work.

Other embeddings of the group manifold to flat space can also be useful. For example, $\mathrm{SU}(2)\simeq\mathrm{S}^3$ can be embedded into $\mathbb{C}^2=\mathbb{R}^4$, interpreting the radial coordinate as a scalar. This approach can cut half of the bosons per link from the orbifold lattice. It would be nice if we could find the best embedding for each group, balancing the qubit requirement and gate complexity.

Our findings open numerous directions for future exploration. We can extend our numerical demonstration to $(3+1)$-dimensional theory, incorporate quarks, increase to $N>3$, and investigate the continuum limit. The constraint term that enforces $\det U = 1$ may prove less critical than previously thought, at least for pure Yang-Mills theory, and verifying this numerically could substantially reduce simulation costs. Additionally, we can estimate the required qubit count for quantum simulations using the Monte Carlo technique introduced in ref.~\cite{Hanada:2022pps}. Developing optimized simulation algorithms --- particularly state preparation methods using orbifold lattice techniques --- represents another critical avenue. In this context, avoiding nontrivial oracles is essential to preserve the advantages offered by the orbifold lattice. Reexamining the Kogut-Susskind Hamiltonian literature through this new lens may reveal valuable insights, such as magnetic basis formulations that could facilitate more efficient state preparations.

The approach presented here represents the first viable method to make the Kogut-Susskind Hamiltonian programmable toward the continuum limit on digital quantum computers for arbitrary gauge groups and dimensions. Our framework uniquely enables seamless utilization of both coordinate (magnetic) and momentum (electric) bases. Furthermore, the orbifold Hamiltonian belongs to a broader class of Hamiltonians \eqref{generic_class}, creating opportunities for developing tailored simulation strategies that fully exploit these capabilities. Near-term goals should include testing simplified models of the same generic form on actual quantum devices as proof-of-principle demonstrations.

The fundamental simplicity of the orbifold lattice stems from emergent geometry, arising from matrix models with specific backgrounds~\cite{Kaplan:2002wv,Arkani-Hamed:2001kyx} -- a concept that extends to noncommutative geometry in matrix models~\cite{Gharibyan:2020bab}. Theories on emergent spaces inherit elegant structures from their original formulations, suggesting that nature's constructions may surpass human design, as dynamically generated spatial dimensions exhibit superior properties compared to artificially constructed lattices~\cite{Gharibyan:2020bab}. Gauge/gravity dualities~\cite{Maldacena:1997re} provide even deeper insights, connecting gravitational geometries with non-gravitational theories.\footnote{
It may be useful to note that Kaplan, Katz, and Unsal studied lattice formulation of super Yang-Mills theory motivated by gauge/gravity duality and invented the orbifold lattice construction~\cite{Kaplan_Unsal_private_communication}.
} Enhanced understanding of emergent geometry could lead to more efficient quantum simulation protocols. Importantly, the universal framework for Hamiltonians of the form \eqref{generic_class} applies across diverse theories --- from simple toy models to matrix models and quantum field theories dual to superstring/M-theory. This unifying perspective reveals that quantum simulations of QCD and superstring/M-theory are fundamentally interconnected; by advancing our understanding of both simultaneously, we stand to gain deeper insights and achieve more significant progress in quantum simulation capabilities. 
%%%%%%%%%%%%%%%%%%%%%%
%%%%%%%%%%%%%%%%%%%%%%
\begin{center}
\section*{Acknowledgment}
\end{center}
%%%%%%%%%%%%%%%%%%%%%%
%%%%%%%%%%%%%%%%%%%%%%
The authors thank Graham Van Goffrier, Shunji Matsuura, Enrico Rinaldi, Andreas Sch\"{a}fer, David Schaich, Jesse Stryker, and Carsten Urbach for discussions and comments. 
The authors thank Andreas Sch\"{a}fer for carefully checking the manuscript and providing valuable feedback.
G.~B.\ is  funded by the Deutsche Forschungsgemeinschaft (DFG) under Grant No.~432299911 and 431842497. M.~H.~thanks the STFC for the support through the consolidated grant ST/Z001072/1. E.~M. was supported by UK Research and Innovation Future Leader Fellowship {MR/X015157/1}.
Part of the computing time for this project has been provided by the compute cluster ARA of the University of Jena. Part of the numerical simulations were undertaken on Barkla High Performance Computing facilities at the University of Liverpool.
The authors acknowledge the use of the AI-based assistant Claude (Anthropic) for editing assistance during the preparation of this manuscript. The authors take full responsibility for all content of the article.
\\
\bibliographystyle{utphys}
\bibliography{references}

\end{document}